\pdfoutput=1
\documentclass[11pt]{article}

\usepackage{natbib}
\usepackage{comment}
\usepackage{fullpage}
\usepackage{graphicx}
\usepackage[font=small]{caption}
\usepackage{subcaption}
\usepackage{amssymb}
\usepackage{amsmath}
\usepackage{graphics}
\usepackage{bm}
\usepackage{color}
\usepackage{amsthm}
\usepackage{enumerate}
\usepackage{amsfonts}
\usepackage{bbm}
\usepackage[toc]{appendix}
\usepackage{enumitem}
\usepackage[affil-it]{authblk}

\usepackage[compact]{titlesec}
\titlespacing{\section}{0pt}{2ex}{1ex}
\titlespacing{\subsection}{0pt}{1ex}{0ex}
\titlespacing{\subsubsection}{0pt}{0.5ex}{0ex}

\usepackage{tikz}
\usepackage[CJKbookmarks=true,
bookmarksnumbered=true,
bookmarksopen=true,
colorlinks=true,
citecolor=blue,
linkcolor=blue,
anchorcolor=blue,
urlcolor=blue]{hyperref}
\usepackage[ruled, vlined, lined, commentsnumbered,linesnumbered]{algorithm2e}

\DeclareMathOperator*{\argmin}{argmin}

\newtheorem{thm}{Theorem}
\newtheorem{definition}{Definition}
\newtheorem{corollary}{Corollary}
\newtheorem{assumption}{Assumption}
\theoremstyle{definition}
\newtheorem{remark}{Remark}
\usepackage{float}

\def\E{\mathbb{E}}
\def\Cov{\mathrm{Cov}}

\def\rank{\mathrm{rank}}
\def\vec{\mathrm{vec}}

\newcommand{\tp}{\intercal}

\newcommand{\bigO}{\ensuremath{\mathop{}\mathopen{}\mathcal{O}\mathopen{}}}

\newcommand{\bigOp}{\bigO_\mathrm{p}}

\newcommand{\Gscr}{{\mathcal{G}}}

\newcommand{\Hscr}{{\mathcal{H}}}
\newcommand{\Uscr}{{\mathcal{U}}}

\newcommand{\Mscr}{{\mathcal{M}}}

\newcommand{\supp}[1]{#1}

\usepackage{relsize}
\newcommand{\mysquare}{\mathrel{\scalebox{0.5}{$\blacksquare$}}}
\newcommand{\squareun}{{\mysquare}}

\newcommand{\kc}{b}
\newcommand{\uni}{c}
\newcommand{\vc}{c_\eta}
\newcommand{\pc}{c_p}

\newcommand{\tex}[1]{#1}

\newcommand{\Wong}{\textsf{OpCov}}
\newcommand{\convex}{\textsf{mOpCov}}
\newcommand{\llsm}{\textsf{ll-smooth}}
\newcommand{\llsmp}{\textsf{ll-smooth+}}

\newcommand{\Cs}{\gamma}
\newcommand{\C}{\Gamma}

\begin{document}
\title{Low-Rank Covariance Function Estimation for Multidimensional Functional Data}

\author[1]{Jiayi Wang}
\author[1]{Raymond K. W. Wong\thanks{The research of Raymond K. W. Wong is partially supported by National Science Foundation grants DMS-1806063, DMS-1711952 and CCF-1934904.
}}
\author[2]{Xiaoke Zhang\thanks{The research of Xiaoke Zhang is partially supported by National Science Foundation grant DMS-1832046.}}
\affil[1]{Department of Statistics, Texas A\&M University}
\affil[2]{Department of Statistics, George Washington University}

\maketitle

\begin{abstract}
  Multidimensional function data arise from many fields nowadays. The covariance function plays an important role in the analysis of such increasingly common data. In this paper, we propose a novel nonparametric covariance function estimation approach under the framework of reproducing kernel Hilbert spaces (RKHS) that can handle both sparse and dense functional data. We extend multilinear rank structures for (finite-dimensional) tensors to functions, which allow for flexible modeling of both covariance operators and marginal structures. The proposed framework can guarantee that the resulting estimator is automatically semi-positive definite, and can incorporate various spectral regularizations. The trace-norm regularization in particular can promote low ranks for both covariance operator and marginal structures. Despite the lack of a closed form, under mild assumptions, the proposed estimator can achieve unified theoretical results that hold for any relative magnitudes between the sample size and the number of observations per sample field, and the rate of convergence reveals the “phase-transition” phenomenon from sparse to dense functional data. Based on a new representer theorem, an ADMM algorithm is developed for the trace-norm regularization. The appealing numerical performance of the proposed estimator is demonstrated by a simulation study and the analysis of a dataset from the Argo project.
\end{abstract}


\section{Introduction}\label{sec:intro}

In recent decades, functional data analysis (FDA) has become a popular
branch of statistical research.
General introductions to FDA can be found in a few monographs
\citep[e.g.,][]{RamsS05, FerrV06, HorvK12, HsinE15, KokoR17}.
While traditional FDA
deals with a sample of time-varying trajectories, many new forms of functional data have emerged
due to improved capabilities of data recording and storage,
as well as advances in scientific computing.
One particular new form of functional data is \emph{multidimensional functional
	data}, which
becomes increasingly common in various
fields such as climate science, neuroscience and chemometrics. Multidimensional functional data are
generated from
random fields, i.e., random functions of several $\textit{input}$ variables.
One example is multi-subject magnetic resonance imaging (MRI) scans, such as those collected by the Alzheimer's Disease Neuroimaging Initiative.
A human brain is virtually divided into three-dimensional boxes called ``voxels'' and brain signals obtained from these voxels form a three-dimensional functional sample indexed by spatial locations of the voxels.
Despite the growing popularity of multidimensional functional data, statistical methods for such data are limited apart from very few existing works
\citep[e.g.,][]{,Huang-Shen-Buja09, allen2013multi, Zhang-Shen-Huang13, zhou2014principal, wang2017regularized}.

In FDA covariance function estimation plays an important role.
Many methods have been proposed for unidimensional functional data
\citep[e.g.,][]{rice1991estimating, james2000principal,yao2005functionala,
	paul2009consistency, 	li2010uniform, goldsmith2011penalized, xiao2013fast},
and a few were particularly developed for two-dimensional functional data \citep[e.g.,][]{zhou2014principal, wang2017regularized}. In general when the input domain is of dimension $p$,
one needs to estimate a
$2p$-dimensional covariance function. Since covariance function estimation in
FDA is typically nonparametric, the curse of dimensionality emerges soon when $p$ is
moderate or large.

For general $p$,
most work are restricted to
regular and fixed designs \citep[e.g.,][]{zipunnikov2011multilevel, allen2013multi}, where all random fields
are observed over a regular grid like MRI scans.
Such sampling plan leads to a tensor dataset, so
one may apply tensor/matrix decompositions to estimate the covariance function.
When random fields are observed at irregular
locations, the dataset is no longer
a completely observed tensor so tensor/matrix decompositions are not directly applicable.
If observations are densely
collected for each random field,
a two-step approach is a natural solution,
which involves pre-smoothing every random field
followed by ensor/matrix decompositions at a fine discretized grid.
However, this solution is infeasible for sparse data where there are a limited number of observations per random field. One example is the data collected by the international Argo project (\url{http://www.argo.net}). See Section \ref{sec:real} for more details.
In such sparse data setting, one may apply the local smoothing method of \citet{Chen-Jiang17},
but it suffers from the curse of dimensionality when the dimension $p$ is moderate due to a $2p$-dimensional nonparametric regression.

We notice that there is a related class of literature
on longitudinal functional
data \citep[e.g.,][]{Chen-Muller12, park2015longitudinal, Chen-Delicado-Muller17},
a special type of multidimensional functional data where a function is repeatedly
measured over longitudinal times. Typically multi-step methods are needed to
model the functional and longitudinal dimensions either separately (one
dimension at a time) or sequentially (one dimension given the other), as opposed
to the joint estimation procedure proposed in this paper. We also notice
a recent work on longitudinal functional
data under the Bayesian framework \citep{Shamshoian-Senturk-Jeste19}.

The contribution of this paper is three-fold.
First, we propose a new and flexible nonparametric method
for low-rank covariance function estimation for multidimensional functional data,
via the introduction of (infinite-dimensional) unfolding operators (See Section \ref{sec:unfold}).
This method can handle both sparse and dense functional data,  and 
can achieve joint 
structural reductions in all dimensions,
in addition to rank reduction of the covariance operator.
The proposed estimator is
guaranteed to be semi-positive definite.
As a one-step procedure,
our method reduces the theoretical complexities compared to multi-steps estimators which often involve a functional principal component analysis followed by a truncation and reconstruction step
\citep[e.g.,][]{Hall-Vial06,Poskitt-Sengarapillai13}.

Second, we generalize the representer theorem  for unidimensional
functional data by \cite{Wong-Zhang19} to the multidimensional case with more complex spectral
regularizations. The new representer theorem makes the estimation procedure
practically computable
by generating a finite-dimensional parametrization
to the solution of the underlying infinite-dimensional optimization.

Finally, a unified asymptotic theory is developed for the proposed estimator.
It automatically incorporates the settings of dense and sparse functional data, and
reveals a phase transition
in the rate of convergence.
Different from existing theoretical work
heavily based on closed-form representations of estimators, \citep{li2010uniform, Cai-Yuan10, zhang2016sparse, liebl2019inference},
this paper provides the first unified theory for
penalized global M-estimators of covariance functions
which
does not require a closed-form solution. Furthermore, a near-optimal (i.e.,
optimal up to a logarithmic order)
one-dimensional nonparametric rate of convergence is attainable for the
$2p$-dimensional covariance function estimator for
Sobolev-Hilbert spaces.

The rest of the paper is organized as follows.  Section \ref{sec:setup} provides
some background on
reproducing kernel Hilbert space
(RKHS) frameworks for functional data. Section \ref{sec:unfold} introduces
Tucker decomposition for finite-dimensional tensors and our proposed
generalization to tensor product RKHS operators, which is the foundation for our
estimation procedure. The proposed estimation method is given in Section
\ref{sec:estimate}, together with an computational algorithm.
The unified theoretical results are presented in Section \ref{sec:theory}.  The numerical performance of the proposed method is evaluated by a simulation study in Section \ref{sec:simulation} and a real data application in Section \ref{sec:real}.

 \section{RKHS Framework for Functional Data}\label{sec:setup}
In recent years there is a surge of RKHS methods in FDA
\citep[e.g.,][]{Yuan-Cai10, Zhu-Yao-Zhang14, Li-Song17, Reimherr-Sriperumbudur-Taoufik18, Sun-Du-Wang18, Wong-Li-Zhu19}. However, covariance function estimation,
a seemingly well-studied problem,
does not receive the same amount of attention in the development of RKHS methods, even for unidimensional functional data.
Interestingly, we find that the RKHS modeling provides a versatile framework for both unidimensional and multidimensional functional data.

Let $X$ be a random field defined on an index set $\mathcal{T}\subset \mathbb{R}^p$,
with a mean function $\mu_0(\cdot) = \E\{X(\cdot)\}$ and a covariance function $
\Cs_0(*, \cdot) = \Cov(X(*), X(\cdot))$, and let $\{X_i: i=1, \ldots, n\}$ be $n$ independently and identically distributed (i.i.d.)~copies of $X$. Typically, a functional dataset is represented by
$\{(\bm T_{ij},Y_{ij}):  j=1,\dots m_i; i=1,\dots, n\}$,
where
\begin{equation}
Y_{ij} = X_i(\bm T_{ij}) +\epsilon_{ij} \in \mathbb{R}
\label{eqn:obsmodel}
\end{equation}
is the noisy measurement of the $i$-th random field $X_i$ taken at the corresponding index $\bm T_{ij}\in \mathcal{T}$, $m_i$ is the number of measurements observed from the $i$-th random field, and $\{\epsilon_{ij}: i=1,\dots, n; j=1,\dots m_i\}$ are independent~errors with mean zero and finite variance.
For simplicity and without loss of generality, we assume $m_{i}=m$ for all $i$.

As
in many nonparametric regression setups such as
penalized regression splines \citep[e.g.,][]{Pearce-Wand06} and smoothing splines
\citep[e.g.,][]{Wahba90, Gu13}, the sample field of $X$, i.e., the realized $X$
(as opposed to the sample path of a unidimensional random function),
is assumed to reside in
an RKHS $\mathcal{H}$ of
functions defined on $\mathcal{T}$
with a continuous and square integrable reproducing kernel $K$.
Let $\langle \cdot, \cdot\rangle_{\mathcal{H}}$ and $\|\cdot\|_{\Hscr}$ denote the inner product and norm of $\mathcal{H}$ respectively.
With the technical condition $\E\|X\|^2_{\mathcal{H}}<\infty$,
the covariance function
$\Cs_0$ resides in the tensor product RKHS $\mathcal{H}\otimes \Hscr$.
It can be shown that $\Hscr\otimes \Hscr$
is an RKHS, equipped with the reproducing kernel $K \otimes K$ defined as
$(K \otimes K) ((\bm s_1, \bm t_1), (\bm s_2, \bm t_2)) = K(\bm s_1, \bm s_2) K(\bm t_1, \bm t_2)$,
for any $\bm s_1,\bm s_2,\bm t_1,\bm t_2\in\mathcal{T}$.
This result has been exploited by \citet{Cai-Yuan10} and \citet{Wong-Zhang19}
for covariance estimation
in the unidimensional setting.

For any function $f\in\mathcal{H}\otimes \mathcal{H}$, there exists an operator mapping
$\mathcal{H}$ to $\mathcal{H}$ 	defined by
$g\in\mathcal{H}\mapsto \langle f(*,\cdot) , g(\cdot)\rangle_{\mathcal{H}} \in\mathcal{H}$.
When $f$ is a covariance function, we call the induced operator
a $\mathcal{H}$-covariance operator, or simply a covariance operator as below.
To avoid clutter, the induced operator will share the same notation with the generating function.
Similar to $L^2$-covariance operators,
the definition of an induced operator is obtained by replacing the $L^2$ inner product
by the RKHS inner product.
The benefits of considering this operator have been discussed in \cite{Wong-Zhang19}.
We also note that a singular value decomposition \citep[e.g.,][]{HsinE15}
of the induced operator exists whenever the corresponding function $f$ belongs to the tensor product RKHS
$\mathcal{H}\otimes \mathcal{H}$.
{The idea of induced operator can be similarly extended to general tensor product
	space $\mathcal{F}_1\otimes \mathcal{F}_2$ where $\mathcal{F}_1$ and
	$\mathcal{F}_2$ are two generic RKHSs of functions.}

For any $\Cs \in \mathcal{H}\otimes \mathcal{H}$, let $\Cs^\top$ be the transpose of $\Cs$, i.e., $\Cs^\top(\bm s, \bm t)=\Cs(\bm t, \bm s)$, $\bm s, \bm t \in \mathcal{T}$. Define
$\Mscr=\{\Cs\in\Hscr\otimes \Hscr: \Cs\equiv \Cs^\top\}$.
To guarantee symmetry and positive semi-definiteness of the
estimators,
\citet{Wong-Zhang19}
adopted
$\Mscr^+=\{ \Cs\in\Mscr: \langle \Cs f, f\rangle_{\Hscr} \ge 0, \forall f\in\Hscr\}$ as the hypothesis class of $\Cs_0$
and considered the following regularized estimator:
\begin{equation}
\underset{\Cs\in\Mscr^+}{\arg\min} \left\{ \ell(\Cs) + \tau \Psi(\Cs) \right\},
\label{eqn:ggoal}
\end{equation}
where
$\ell$ is a convex and smooth loss function characterizing the fidelity to the data,
$\Psi(\Cs)$ is a spectral penalty function
(see Definition \ref{def:specfun} below), and $\tau$ is a tuning parameter.
Due to the constraints specified in
$\Mscr^+$, the resulting covariance estimator
is always positive semi-definite.
In particular, if the spectral penalty function $\Psi(\Cs)$ imposes the trace-norm
regularization, an $\ell_1$-type shrinkage penalty on the respective singular values,
the estimator
is usually of low rank.
\citet{Cai-Yuan10} adopted a similar objective function 
as in (\ref{eqn:ggoal}) but with the hypothesis class $\Hscr\otimes \Hscr$ and
an $\ell_2$-type penalty $\Psi(\Cs) = \|\Cs\|_{\Hscr\otimes \Hscr}^2$,
so the resulting estimator may neither be positive semi-definite nor low-rank.

Although \citet{Cai-Yuan10} and \citet{Wong-Zhang19} focused
on unidimensional functional data,
their frameworks can be directly extended to the multidimensional setting.
Explicitly, similar to (\ref{eqn:ggoal}), as long as a proper $\Hscr$ for the random fields with dimension $p > 1$ is selected, an efficient ``one-step'' covariance function estimation with the hypothesis class $\Mscr^+$ can be obtained immediately, which results in a positive semi-definite and possibly low-rank estimator.
Since an RKHS is identified by its reproducing kernel,
we simply need to pick a multivariate reproducing kernel $K$
for multidimensional functional data.
However, even when the low-rank approximation/estimation
is adopted (e.g., by trace-norm regularization),
we still need to estimate several $p$-dimensional eigenfunctions nonparametrically.
This curse of dimensionality calls for
a more efficient modeling.
Below, we explore this through the lens of
tensor decomposition in finite-dimensional vector spaces
and
its extension to
infinite-dimensional function spaces.

\section{Low-Rank Modeling via Functional Unfolding}\label{sec:unfold}

In this section we will extend the well-known Tucker decomposition for finite-dimensional tensors to functional data, then introduce the concept of functional unfolding for low-rank modeling, and finally apply functional unfolding to covariance function estimation.

\subsection{Tucker decomposition for finite-dimensional tensors}
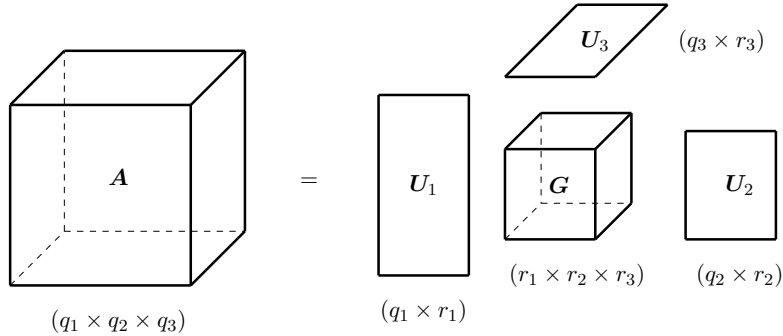
\begin{figure}[ht]
	\centering
	\scalebox{0.8}{
		\begin{tikzpicture}[baseline=10ex,scale=3]
		\coordinate (A1) at (0, 0);
		\coordinate (A2) at (0, 1);
		\coordinate (A3) at (1, 1);
		\coordinate (A4) at (1, 0);
		\coordinate (B1) at (0.3, 0.3);
		\coordinate (B2) at (0.3, 1.3);
		\coordinate (B3) at (1.3, 1.3);
		\coordinate (B4) at (1.3, 0.3);
		\draw[very thick] (A1) -- (A2);
		\draw[very thick] (A2) -- (A3);
		\draw[very thick] (A3) -- (A4);
		\draw[very thick] (A4) -- (A1);
		\draw (0.6,0.6) node{$\bm A$};
		\draw (0.6,-0.2) node{$(q_1\times q_2 \times q_3)$};
		\draw[dashed] (A1) -- (B1);
		\draw[dashed] (B1) -- (B2);
		\draw[very thick] (A2) -- (B2);
		\draw[very thick] (B2) -- (B3);
		\draw[very thick] (A3) -- (B3);
		\draw[very thick] (A4) -- (B4);
		\draw[very thick] (B4) -- (B3);
		\draw[dashed] (B1) -- (B4);
		\end{tikzpicture}
		\qquad
		=
		\qquad
		\begin{tikzpicture}[baseline=9ex,scale=3]
		\draw (0.25,0.5) node{$\bm U_1$};
		\draw (0.25,-0.2) node{$(q_1\times r_1)$};
		\draw[very thick] (0,0) -- (0.5,0);
		\draw[very thick] (0.5,0) -- (0.5,1);
		\draw[very thick] (0.5,1) -- (0,1);
		\draw[very thick] (0,1) -- (0,0);
		\draw[very thick] (0.7,0.2)--(1.2,0.2);
		\draw[very thick] (1.2,0.2) -- (1.2,0.7);
		\draw[very thick](1.2,0.7)--(0.7,0.7);
		\draw[very thick] (0.7,0.7)--(0.7,0.2);
		\draw (1,0.5) node{$\bm G$};
		\draw (1.1,0) node{$(r_1\times r_2 \times r_3)$};
		\draw[dashed] (0.7,0.2) -- (0.9,0.4);
		\draw[very thick] (1.2,0.2) -- (1.4,0.4);
		\draw[very thick] (1.4,0.4) -- (1.4,0.9);
		\draw[very thick] (1.4,0.9) -- (1.2,0.7);
		\draw[very thick] (1.4,0.9) -- (0.9,0.9);
		\draw[very thick] (0.9,0.9) -- (0.7,0.7);
		\draw[dashed] (0.9,0.9) -- (0.9,0.4);
		\draw[dashed] (0.9,0.4) -- (1.4,0.4);
		\draw(1.2,1.3) node {$\bm U_3$};
		\draw(1.9,1.3) node {$(q_3 \times r_3)$};
		\draw[very thick] (0.7,1.1) -- (1.2,1.1);
		\draw[very thick] (1.2,1.1) -- (1.6,1.5);
		\draw[very thick] (1.6,1.5) -- (1.1,1.5);
		\draw[very thick] (1.1,1.5) -- (0.7,1.1);
		\draw (2.0,0.5) node{$\bm U_2$};
		\draw (2.0,0) node{$(q_2 \times r_2)$};
		\draw[very thick] (1.7,0.2) -- (2.2,0.2);
		\draw[very thick] (2.2,0.2) -- (2.2,0.8);
		\draw[very thick] (2.2,0.8) -- (1.7,0.8);
		\draw[very thick] (1.7,0.8) -- (1.7,0.2);
		\end{tikzpicture}
	}
	\caption{Tucker decomposition of a third-order array. The values in the parentheses are dimensions for the corresponding matrices or arrays.}
	\label{tucker_figure}
\end{figure}

First, we give a brief introduction to the popular Tucker decomposition \citep{tucker1966some}
for \textit{finite-dimensional} tensors.
Let $\mathcal{G}= \bigotimes_{k=1}^{d} \Gscr_k$ denote a generic tensor product space with finite-dimensional $\mathcal{G}_k, k=1, \ldots, d$.
If the dimension of $\mathcal{G}_k$ is $q_k$, $k=1, \ldots, d$, then each element in $\mathcal{G}= \bigotimes_{k=1}^{d} \Gscr_k$ can be identified by
an array in $\mathbb{R}^{\prod_{j=k}^d q_k}$, which contains the coefficients through an orthonormal basis.
By Tucker decomposition, any array in $\mathbb{R}^{\prod^d_{k=1} q_k}$ can be represented in terms of $n$-mode products
as follows.
\begin{definition}[$n$-mode product]
	\label{def:n-mode}
	For any arrays $\bm A\in\mathbb{R}^{q_1\times q_2\times\cdots\times q_d}$ and $\bm P\in  \mathbb{R}^{p_n \times q_n}$,
	$n\in\{1,\dots,d\}$, the $n$-mode product between $\bm A$ and $\bm P$, denoted by $\bm A \times_n \bm{P}$, is a array of dimension \\
	${q_1\times q_2\times\cdots q_{n-1}\times p_n\times q_{n+1}\times\cdots q_d}$
	of which $(l_1,\dots,l_{n-1},j, l_{n+1}, \dots l_d)$-th element is defined by
	$$(\bm A \times_n \bm P)_{l_1,\dots,l_{n-1},j,l_{n+1}, \dots l_d} = \sum_{i=1}^{q_n} \bm A_{l_1,\dots,l_{n-1},i ,l_{n+1}, \dots l_d} \bm P_{j, i}.
	$$
\end{definition}
\begin{definition}[Tucker decomposition]
	\label{def:tucker}
	Tucker decomposition of
	$\bm A\in\mathbb{R}^{q_1\times q_2\times\cdots\times q_d}$
	is
	\begin{equation}
	\label{tucker_form}
	\bm A = \bm G \times_1 \bm U_1 \times_2\cdots\times_d \bm U_d,
	\end{equation}
	where $\bm U_i \in \mathbb{R}^{q_i \times r_i}$ $i=1,2,\dots,d$, are
	called the ``factor matrices'' (usually orthonormal) with $r_i \leq q_i$
	and $\bm G \in \mathbb{R}^{r_1\times \cdots\times r_d}$ is called the ``core tensor''.
\end{definition}
Figure \ref{tucker_figure} provides a pictorial illustration of a Tucker decomposition.
Unlike matrices, the concept of rank is more complicated for
arrays of order 3 or above.
Tucker decomposition naturally leads to a particular form of rank, called ``multilinear rank'',
which is directly related to the familiar concept of
matrix ranks.
To see this, we employ a reshaping operation called \emph{matricization},
which
rearranges elements of an array into a matrix.
\begin{definition}[Matricization]
	For any $n\in\{1,\dots,d\}$,
	the $n$-mode matricization of $\bm{A}\in\mathbb{R}^{q_1\times q_2\times\cdots\times q_d}$, denoted by $\bm{A}_{(n)}$,
	is a matrix of dimension $q_n\times (\prod_{k\neq n} q_k)$
	of which $(l_n,j)$-th element is defined by
	$[\bm A_{(n)}] _{l_n, j}=\bm A_{l_1,\ldots, l_d}$, where $j = 1 +
	\sum_{i=1,i\neq n}^d (l_i-1)(\prod_{m=1, m\neq n}^{i-1}q_m)$\footnote{All
		empty products are defined as 1. For example, $\prod_{m=i}^{j}q_m = 1$ when $i>j$.}.
\end{definition}
For any $\bm A\in\mathbb{R}^{q_1\times q_2\times\cdots\times q_d}$, by simple derivations, one can obtain a useful
relationship between the $n$-mode matricization and Tucker decomposition
$\bm A = \bm G \times_1 \bm U_1 \times_2\cdots\times_d \bm U_d$:
\begin{equation}
\bm{A}_{(n)} = \bm{U}_n \bm{G}_{(n)} ( \bm{U}_d \otimes \cdots \otimes \bm{U}_{n+1}\otimes \bm{U}_{n-1}\otimes \cdots\otimes \bm{U}_1 )^\tp,
\label{eqn:n-mode-tucker}
\end{equation}
where, with a slight abuse of notation, $\otimes$ also represents the Kronecker product between matrices.
Hence if the factor matrices are of full rank, then
$\rank(\bm{A}_{(n)}) = \rank(\bm{G}_{(n)})$.
The vector $(\rank(\bm{A}_{(1)}), \dots, \rank(\bm{A}_{(d)}))$ is
known as the \emph{multilinear rank} of $\bm{A}$.
Clearly from \eqref{eqn:n-mode-tucker},
one can choose a Tucker decomposition such that
$\{\bm{U}_k: k=1,\ldots, d\}$ are orthonormal matrices
and $\mathrm{rank}(\bm{U}_k)=r_k$.
Therefore a ``small" multilinear rank corresponds to a small core tensor
and thus an intrinsic dimension reduction, which potentially improves estimation and interpretation.
We will relate this low-rank structure to
multidimensional functional data.
\subsection{Functional unfolding for infinite-dimensional tensors}
\label{fc_unfold}
To encourage low-rank structures in covariance function estimation,
we generalize the matricization operation for finite-dimensional arrays to infinite-dimensional tensors \citep{Hackbusch12}.
Here let $\mathcal{G}= \bigotimes_{k=1}^{d} \Gscr_k$ denote a generic tensor product space
where $\mathcal{G}_k$ is an RKHS of functions with an inner product  $\langle \cdot,  \cdot \rangle_{\Gscr_k}$,
for $k=1, \ldots, d$.

Notice that the tensor product space $\Gscr = \bigotimes_{k=1}^d \Gscr_k$
can be generated by some elementary tensors
of the form $\bigotimes^d_{k=1} f_k(x_1,\dots,x_d) = \prod^d_{k=1}f_k(x_k)$ where $f_k\in\Gscr_k, k=1, \ldots, d$.
More specifically, $\mathcal{G}$ is the completion of the linear span
of all elementary tensors under the inner product
$\langle \bigotimes^d_{k=1} f_k, \bigotimes^d_{k=1} f'_k \rangle_\Gscr = \prod^d_{k=1} \langle f_k, f_k'\rangle_{\Gscr_k}$, for any $f_k, f_k'\in\Gscr_k$.

In Definition \ref{def:unfolding} below,
we generalize matricization/unfolding for finite-dimensional arrays to infinite-dimensional elementary tensors. We also define a square unfolding for infinite-dimensional tensors that will be used
to describe the spectrum of covariance operators.

\begin{definition}[Functional unfolding operators]\label{def:unfolding}
	The one-way unfolding operator and square unfolding operators are defined as follows for
	any elementary tensor of the form
	$\bigotimes^d_{k=1}f_k$.
	\begin{enumerate}
		\item One-way unfolding operator $\Uscr_j$ for $j=1,\dots,d$:
		The $j$-mode one-way unfolding operator
		$\Uscr_j: \bigotimes_{k=1}^d  \mathcal{G}_k \rightarrow   \mathcal{G}_j \otimes  (\bigotimes_{k\neq j} \mathcal{G}_k)$
		is defined by
		$\Uscr_j(\bigotimes_{k=1}^d f_{k}) 
		= f_j \otimes ( \bigotimes_{k\neq j} f_k )$.
		\item Square unfolding operator $\mathcal{S}$:
		When $d$ is even,
		the square unfolding operator
		$\mathcal{S}: \bigotimes_{j=1}^{d}  \mathcal{G}_j \rightarrow  ( \bigotimes^{d/2}_{j=1} \mathcal{G}_j ) \otimes  ( \bigotimes^{d}_{k=d/2+1} \mathcal{G}_k )$
		is defined by
		$\mathcal{S}(\bigotimes^d_{j=1} f_{j}) =
		( \bigotimes_{j=1}^{d/2}f_j )
		\otimes
		( \bigotimes_{k=d/2+1}^{d}f_k )$.
	\end{enumerate}
	These definitions extend to any function $f\in \mathcal{G}$ 
	by linearity. For notational simplicity we denote $\Uscr_j(f)$ by $f_{(j)}$, $j=1,\ldots, d$, and $\mathcal{S}(f)$ by $f_\squareun$.
\end{definition}

Note that the range of each functional unfolding operator, either $\Uscr_j$, $j=1,\ldots, d$ or $\mathcal{S}$, is a tensor product of \textit{two} RKHSs, so its output can be interpreted as an (induced)
operator. Given a function $f\in\mathcal{G}$, the multilinear rank
can be defined as $(\rank(f_{(1)}),\dots, \rank(f_{(d)}))$,
where $f_{(j)}$'s are interpreted as an operator here and
$\mathrm{rank}(A)$ is the rank of any operator $A$. If all $\mathcal{G}_k, k=1, \ldots, d$ are finite-dimensional, the singular values of the output of any functional unfolding operator
match with those of the $j$-mode matricization (of the corresponding array representation).

\subsection{Functional unfolding for covariance functions}
\label{fc_unfold_cov}

Suppose that the random field $X \in \Hscr = \bigotimes_{k=1}^p \Hscr_k$ where each $\Hscr_k$ is a RKHS of functions equipped with an inner product $\langle \cdot,*\rangle_{k}$ and corresponding
norm $\|\cdot\|_k$, $k=1,\ldots, p$. Then the covariance function $\Cs_0$ resides in
$\Hscr\otimes \Hscr=
(\bigotimes_{j=1}^p \Hscr_j) \otimes (\bigotimes_{k=1}^p \Hscr_k)$.
To estimate $\Cs_0$, we could consider a special case of $\mathcal{G}=\bigotimes^d_{j=1} \mathcal{G}_j$ in Section \ref{fc_unfold} by letting $d=2p$, $\mathcal{G}_j=\mathcal{H}_j$ for $j=1,\dots,p$;
$\mathcal{G}_j =\mathcal{H}_{j-p}$ for $j=p+1,\dots,d$; and $\langle
\cdot,\cdot\rangle_{\Gscr_j} = \langle \cdot,\cdot\rangle_{j}$ for $j=1, \ldots, d$.

Clearly, the elements of $\Hscr \otimes \Hscr$
are identified by those in
$\mathcal{G}=\bigotimes^d_{j=1} \mathcal{G}_j$.
In terms of the folding structure, $\Hscr\otimes \Hscr$
has a squarely unfolded structure.
Since a low-multilinear-rank structure is represented by
different unfolded forms, it would be easier to study the completely folded space $\bigotimes^{d}_{k=1}\Gscr_k$ instead of the squarely unfolded space $\Hscr\otimes \Hscr$.
We use $\C_0$ to represent the folded covariance function, the corresponding element of $\Cs_0$ in $\Gscr$.
In other words, $\C_{0,\mysquare}=\Cs_0$.
For any $\C\in\Gscr$, $\mathrm{rank}(\C_\squareun)$ is defined as the \textit{two-way rank} of $\C$ while $\mathrm{rank}(\C_{(1)}), \dots, \mathrm{rank}(\C_{(p)})$ are defined as the \textit{one-way ranks} of $\C$.

\begin{remark}\label{rmk:unfolding}
	For an array $\bm{A}\in\mathbb{R}^{\prod_{k=1}^dq_k}$,
	the one-way unfolding $\Uscr_j(\bm A)$
	is the same as matricization,
	if we further impose the same ordering of the columns in the output of $\mathcal{U}_j(\bm{A})$, $j=1, \ldots, d$.
	This ordering is just related to how we represent the array, and is not crucial in the general definition of $\mathcal{U}_j$. Since the description of the computation strategy depends on the explicit representation,
	we will always assume this ordering.
	Similarly,
	we also define a specific ordering of rows and columns
	for ${\bm A}_\squareun\in\mathbb{R}^{(d/2)\times (d/2)}$ when $d$ is even,
	such that its $(j_1, j_2)$-th entry is $\bm{A}_{k_1,\dots,k_d}$
	where $j_{1} = 1 + \sum_{i=1}^{d/2} (k_i-1)(\prod_{m=i+1}^{d/2} q_m)$ and $j_{2} = 1 + \sum_{i=d/2+1}^{d} (k_i-1)(\prod_{m=i+1}^{d} q_m)$. 
\end{remark}

\subsection{One-way and two-way ranks in covariance functions}
\label{sec:lowrankcov}
Here we illustrate the roles of one-way and two-way ranks
in the modeling of covariance functions.
For a general $\mathcal{G}=\bigotimes^d_{j=1} \mathcal{G}_j$,
let $\{e_{k,l_k}:l_j=1,\dots,q_k\}$ be a set of orthonormal basis functions
of
$\Gscr_k$ for $k=1,\dots, d=2p$,
where $q_k$ is allowed to be infinite, depending on the dimensionality of $\Gscr_k$.
Then $\{\bigotimes^d_{k=1}e_{k,l_k}: l_k=1,\dots,q_k; k=1,\dots, d \}$
forms a set of orthonormal basis functions for $\Gscr$.
Thus for any $\C\in\Gscr$, we can express
\begin{align}\label{eq:cov}
\C &= \sum_{k_1, k_2,\dots, k_d} \bm{B}_{k_1,\dots, k_d} \bigotimes_{i=1}^{d} e_{i,k_i},
\end{align}
where the coefficients $\bm{B}_{k_1,\dots, k_d}$ are real numbers.
For convenience, we collectively put them into
an array $\bm{B}\in\mathbb{R}^{\prod^d_{k=1}q_k}$.

To illustrate the low-multilinear-rank structures for covariance functions, we consider $p=2$, i.e., $d=2p=4$, and then
by (\ref{eq:cov})
the folded covariance function $\C$ can be expressed by
$$\C(s_1,s_2,t_1,t_2)=\sum_{k_1= 1}^{q_1}\sum_{k_2=1}^{q_2}\sum_{k_3= 1}^{q_1}\sum_{k_4= 1}^{q_2}\bm B_{k_1,k_2,k_3,k_4}e_{1,k_1}(s_1)e_{2,k_2}(s_2)e_{1,k_3}(t_1)e_{2,k_4}(t_2).
$$
To be precise, the covariance function is the squarely unfolded $\C_{\squareun}((s_1,s_2), (t_1,t_2)) \equiv \C(s_1,s_2,t_1,t_2)$.
Suppose that $\bm B$ possesses (or is well-approximated by) a structure of a low multilinear rank,
and yields Tucker decomposition $\bm B = \bm E \times_1 \bm U_1\times_2 \bm U_2  \times_3 \bm U_1\times_{4} \bm U_2$\footnote{Definition \ref{def:n-mode} is extended to the case when $q_n$ is infinite.}
where $\bm E\in\mathbb{R}^{r_1\times r_2\times r_1\times r_2}$, $\bm U_k\in \mathbb{R}^{q_k\times r_k}$ for $k=1,2$, and columns of $\bm U_k$ are orthonormal.
Apparently $R:=\rank(\bm B_\squareun)$ is the two-way rank of $\C$, while
$r_1$ and $r_2$ are the corresponding one-way ranks.
Now the covariance function can be further represented as
$$\C(s_1,s_2,t_1,t_2)=\sum_{j_1=1}^{r_1}\sum_{j_2=1}^{r_2}\sum_{j_3=1}^{r_1}\sum_{j_4=1}^{r_2}\bm E_{j_1,j_2,j_3,j_4}u_{j_1}(s_1)v_{j_2}(s_2)u_{j_3}(t_1)v_{j_4}(t_2),$$
where
$\{u_{j}: j=1, \ldots, r_1\}$ and $\{v_{k}: k=1, \ldots, r_2\}$ are (possibly infinite) linear combinations of the original basis functions.
In fact, $\{u_{j}: j=1, \ldots, r_1\}$ and $\{v_{k}: k=1, \ldots, r_2\}$ are the sets of orthonormal functions of $\Gscr_1$ and $\Gscr_2$ respectively.
Apparently $\mathrm{rank}(\bm{E}_{\squareun})=R$.

Consider the eigen-decomposition of the squarely unfolded core tensor $\bm E_\squareun = \bm P\bm D\bm P^T$ where $\bm{D}=\mathrm{diag}(\lambda_1,\lambda_2,\dots,\lambda_R)$
and $\bm{P}\in\mathbb{R}^{r_1r_2 \times R}$ has orthonormal columns.
Then we obtain the eigen-decomposition of the covariance function $\C_\squareun$:
$$\C_\squareun((s_1,s_2),(t_1,t_2))=\sum_{g=1}^R \lambda_g
f_g(s_1,s_2) f_g(t_1,t_2),
$$
where the eigenfunction is
$$
f_g(s_1,s_2)= \sum_{j_1=1}^{r_1}\sum_{j_2=1}^{r_2}\bm P_{j_2+ (j_1-1)r_1,g}u_{j_1}(s_1)v_{j_2}(s_2) =:
\,
\begin{cases}
\sum^{r_1}_{j_1=1} a_{j_1,g}(s_2) u_{j_1}(s_1)\\
\sum^{r_2}_{j_2=1} b_{j_2,g}(s_1) v_{j_2}(s_2)
\end{cases},
$$
with $a_{j_1,g}(\cdot ) = \sum_{j_2=1}^{r_2} \bm P_{j_2+ (j_1-1)r_1,g}v_{j_2}(\cdot) $ and $b_{j_2,g}(\cdot ) = \sum_{j_1=1}^{r_1} \bm P_{j_2+ (j_1-1)r_1,g}u_{j_1}(\cdot) $.

First, this indicates that the two-way rank $R$ is
the same as the rank of the covariance operator.
Second, this shows that
$\{u_{j_1}:j_1=1, \ldots, r_1\}$ is the common basis for the variation along the dimension $s_1$, hence describing the marginal structure along $s_1$. Similarly $\{v_{j_2}: j_2 =1, \ldots, r_2 \}$ is the common basis that characterizes the marginal variation along the dimension $s_2$. We call them the \emph{marginal basis} along the respective dimension. Therefore,  the one-way ranks
$r_1$ and $r_2$ are the minimal numbers of the one-dimensional functions for the dimensions $s_1$ and $s_2$ respectively
that construct all the eigenfunctions of covariance function $\C$.

Similarly, for $p$-dimensional functional data, each eigenfunction
can be represented by a linear combination of $p$-products of univariate functions.
One can then show that the two-way rank $R$ is the same as the rank of the covariance operator and the one-way ranks $r_1,\dots, r_p$ are
the minimal numbers of one-dimensional functions along respective dimensions that
characterize all eigenfunctions of the covariance operator.

\begin{remark}\label{rmk:rank}
	Obviously, $ R\le \prod^p_{k=1} r_k$ for $p$-dimensional functional data.
	If the random field $X$ has the property of ``weak separability" as defined by \citet{lynch2018test}, then $\max(r_1,\dots,r_p) \le R$ so the low-rank structure in terms of $R$ will be automatically translated to low one-way ranks.
	{Note that the construction of our estimator and corresponding theoretical analysis \textit{do not} require separability conditions.}
\end{remark}

Compared to typical low-rank covariance modelings only in terms of $R$,
we also intend to regularize the one-way ranks $r_1,\dots,r_p$ for two reasons.
First, the illustration above shows that the structure of low one-way ranks encourages a
``sharing'' structure of one-dimensional variations among different
eigenfunctions.
Promoting low one-way ranks can facilitate additional
dimension reduction and further
alleviates the curse of dimensionality.
Moreover, one-dimensional marginal structures will provide more details of the covariance function structure and thus help with a better
understanding of $p$-dimensional eigenfunctions.

Therefore,
we will utilize both one-way and two-way structures and propose an estimation procedure that regularizes one-way and two-way ranks jointly and flexibly, with the aim of seeking the ``sharing" of marginal structures
while controlling the number of eigen-components simultaneously.

\section{Covariance Function Estimation}\label{sec:estimate}
In this section we propose a low-rank covariance function estimation framework based on functional unfolding operators and spectral regularizations.
Spectral penalty functions
\citep{Abernethy-Bach-Evgeniou09, Wong-Zhang19}
are defined as follows.
\begin{definition}[Spectral penalty function]
	\label{def:specfun}
	Given a compact operator $A$,
	a spectral penalty function takes the form
	$\Psi(A) = \sum_{k\ge1} \psi(\lambda_k(A))$ with the singular values of the operator $A$, $\lambda_1(A)$, $\lambda_2(A)$, $\dots$
	in a descending order of magnitude
	and a non-decreasing
	penalty function $\psi$
	such that $\psi(0)=0$.
\end{definition}

Recall
$\Hscr=\bigotimes^p_{j=1}\Hscr_j$ and
$\Gscr=\bigotimes^{d}_{j=1}\Gscr_j$ where $d=2p$, $\mathcal{G}_j=\mathcal{H}_j$ for $j=1,\dots,p$, and
$\mathcal{G}_j =\mathcal{H}_{j-p}$ for $j=p+1,\dots,d$. Clearly, a covariance operator is self-adjoint
and positive semi-definite.
Therefore we consider the hypothesis space
$\Mscr^+ = \{\C \in \Mscr: \langle \C_\squareun f,f\rangle_{\Hscr} \ge 0, \text{for all } f\in \Hscr\}$,
where
$\Mscr=\{\C\in\Gscr: \mbox{$\C_\squareun$ is self-adjoint}\}$, and propose a general class of covariance function estimators as follows:
\begin{equation}
\label{min}
\underset{ \C\in\Mscr ^{+}}{\arg\min} \left\{ \ell(\C) + \lambda \left[   \beta \Psi_0 (\C_\squareun) + \frac{1-\beta}{p} \sum_{j=1}^p \Psi_j(\C_{(j)})\right] \right\}  ,
\end{equation}
where $\ell$ is a convex and smooth loss function,
$\{\Psi_j: j=1, \ldots, p\}$ are spectral penalty functions,
and $\lambda\ge 0$ , $\beta\in [0,1]$ are tuning parameters. Here $\Psi_0$ penalizes the squarely unfolded operator $\C_\squareun$ while $\Psi_j$ regularizes one-way unfolded operator $\C_{(j)}$ respectively for $j=1,\dots,p$.
The tuning parameter $\beta $ controls the relative degree of regularization between one-way and two-way singular values.
The larger the $\beta$ is, the more penalty is imposed on the two-way singular values
relative to the one-way singular values.
When $\beta = 1$, the penalization is only on the eigenvalues of the covariance operator (i.e., the two-way singular values),
similarly as \cite{Wong-Zhang19}.

To achieve low-rank estimation, we adopt a special form of \eqref{min}:
\begin{equation}
\label{propmin}
\hat \C  =   \underset{ \C\in\Mscr ^{+}}{\arg\min}  \left\{ \ell_{\mathrm{square}}(\C) + \lambda \left[ \beta  \|\C
_\squareun\|_* + \frac{1-\beta}{p} \sum_{j=1}^p \|\C_{(j)}\|_*\right] \right\},
\end{equation}
where $\|\cdot\|_*$ is the sum of singular values, also called trace norm,
and $\ell_{\mathrm{square}}$ is the squared error loss:
\begin{equation}\label{eq:loss}
{\ell}_{\mathrm{square}}(\C) = \frac{1}{nm(m-1)}\sum_{i=1}^{n} \sum_{1\leq j\neq j'\leq m} \{\C(T_{ij1},\dots,T_{ijp}, T_{ij'1},\dots,T_{ij'p}) - Z_{ijj'}\}^2,
\end{equation}
with $Z_{ijj'}  = \{Y_{ij} - \hat \mu (T_{ij1},\dots,T_{ijp})\} \{Y_{ij'} - \hat \mu (T_{ij'1},\dots,T_{ij'p})\} $, $\hat \mu$ as an estimate of the mean function, and $T_{ijk}$ as the $k$-th element of  location vector $\bm T_{ij}$.
Notice that trace-norm regularizations promote low-rankness of the underlying operators,
hence leading to a low-rank estimation in terms of both the one-way and two-way (covariance) ranks.

\subsection{Representer theorem and parametrization}
Before deriving a computational algorithm,
we notice that the optimization \eqref{propmin}
is an
infinite-dimensional optimization which is generally unsolvable.
To overcome this challenge,
we show that the solution to \eqref{propmin} always lies in a known finite-dimensional sub-space given data, hence allowing a finite-dimensional parametrization.
Indeed, we are able to achieve a stronger result in Theorem \ref{thm_re} which holds for the general class of estimators obtained by \eqref{min}.

Let $\mathcal{L}_{n,m} = \left\lbrace T_{ijk}: i=1,\dots,n, j=1,\dots,m, k=1,\dots,p \right\rbrace $.
\begin{thm}[Representer theorem]
	\label{thm_re}
	If the solution set of \eqref{min} is not empty, there always exists a solution $\C$  lying in the space $\mathcal{G}(\mathcal{L}_{n,m}) := \bigotimes_{k=1}^{2p} \mathcal{K}_k $, where $ \mathcal{K}_{p+k} =  \mathcal{K}_k$  and  \\ $ \mathcal{K}_k  = \mathrm{span} \left\lbrace K_k(T_{ijk}) : i=1,\dots,n, j=1,\dots,m \right\rbrace$ for $k=1,\dots,p$.
	The solution takes the form:
	\begin{equation}
	\label{form}
	\C(s_1,\ldots,s_p,t_1,\ldots,t_p) = \bm{A} \times_{1}\bm z_1^\tp (s_1)\times_{2}\bm z_2^\tp (s_2)\cdots\times_{p}\bm z_p^\tp (s_p) \times_{p+1}\bm z_1^\tp (t_1)\cdots\times_{2p}\bm z_p^\tp (t_p),
	\end{equation}
	where
	the $l$-th element  of $\bm z_k(\cdot)\in\mathbb{R}^{mn}$ is $K(T_{ijk},\cdot )$ with
	$l=(i-1)n+j$.
	Also,
	$\bm A$ is a $2p$-th order tensor where the  dimension of each mode is $nm$ and
	$\bm A_\squareun$ is a symmetric matrix.
\end{thm}

\supp{The proof of Theorem \ref{thm_re} is given in Section S1 of
	the supplementary material.}
By Theorem \ref{thm_re},
we can now only focus on covariance function estimators of the form \eqref{form}.
Let $\bm B = \bm A \times_ 1\bm M_1^T \cdots \times_ p \bm M_p^T \times_{p+1} \bm M_1^T \cdots \times_ {2p} \bm M_p^T$, where $\bm M_k$ is a $nm \times q_k$ matrix such that $\bm M_k \bm M_k^T = \bm K_k =\left[K(T_{i_1,j_1,k},T_{i_2,j_2,k})\right]_{1\leq i_1, i_2\leq n, 1\leq j_1,j_2\leq m}$.
With $\bm B$, we can express
\begin{align}
\label{eqn:eval}
\begin{split}
\C(s_1,\ldots,s_p,t_1,\ldots,t_p)  =    \bm B  &\times _1 \{ \bm M_1^+ \bm z_1(s_1)\}^\tp \cdots \times _p \{ \bm M_p^+ \bm z_p(s_p)  \}^\tp \\
& \times_{p+1}  \{\bm M_1^+ \bm z_1(t_1) \}^\tp  \cdots \times _{2p} \{\bm M_p^+\bm z_p(t_p)  \}^\tp ,
\end{split}
\end{align}
where $z_k(\cdot)$ is defined in Theorem \ref{thm_re} and $\bm M_k^+$ is the Moore–Penrose inverse of matrix $\bm M_k$.

The Gram matrix $\bm K_k$ is often approximately low-rank. For computational simplicity, one could adopt $q_k$ to be significantly smaller than $nm$.  Ideally we can obtain the ``best" low-rank approximation  with respect to the Frobenius norm by  eigen-decomposition, but a full eigen-decomposition is computationally expensive.
Instead, randomized algorithms can be used to obtain low-rank approximations
in an efficient manner \citep{Halko-Martinsson-Tropp09}.

One can easily show that the eigenvalues of the operator $\C_\squareun$ are the same as
those of the matrix $ {\bm B}_\squareun$ and that the singular values of the operator $\C_{(j)}$ are the same as those of the matrix $ {\bm B}_{(j)}$.
Therefore, solving \eqref{propmin} is equivalent to solving the following optimization:
\begin{equation}
\label{obj}
\min_{\bm B} \left\{ \tilde{\ell}_{\mathrm{square}} (\bm B )+\lambda \left[ \beta h( {\bm B}_\squareun)+ \frac{1-\beta}{p}\sum _{k=1}^{p}\left\| {\bm B}_{(j)} \right\|_*\right] \right\},
\end{equation}
where $\| \cdot \|_*$ also represents the trace norm of matrices, $h(\bm H)= \left\| \bm H\right\|_*$ if matrix $\bm H$ is positive semi-definite, and $h(\bm H)=\infty$ otherwise, and $\tilde{\ell}_{\mathrm{square}} (\bm B ) = \ell_{\mathrm{square}} (\C)$, where $\C$ is constructed from \eqref{eqn:eval}.

Beyond estimating the covariance function, one may be further interested in the eigen-decomposition of
$\C_{\squareun}$ via the $L^2$ inner product, e.g., to perform functional
principal component analysis in the usual sense.
Due to the finite-dimensional parametrization,
a closed-form expression of $L^2$ eigen-decomposition  can be derived from our estimator without further discretization or approximation.
In addition, we can obtain a similar one-way analysis in terms of the $L_2$ inner product.
We can define a $L^2$ singular value decomposition via the Tucker form and
obtain the  $L^2$  marginal basis.
Details are given in Appendix \ref{app:L2_trans}.

\subsection{Computational algorithm}
We solve \eqref{obj} by the accelerated alternating direction method of multipliers (ADMM) algorithm \citep{kadkhodaie2015accelerated}.
We begin with an alternative form of \eqref{obj}:
\begin{align}\label{eq:obj2}
& \min_{\bm B\in \mathbb{R}^{q_1\times\cdots \times q_{2p}}}  \left\{ \tilde{\ell }_{\mathrm{square}}(\bm B)+\lambda \beta h(\bm{D}_{0,\squareun})+ \lambda \frac{1-\beta}{p}\sum_{k=1}^p \left\| \bm{D}_{j,(j)} \right\|_*\right\}.\\
& \text{subject to}\  \   \
\bm B = \bm D_0 = \bm D_1 = \cdots = \bm D_p
\end{align}
where $q_{p+k} = q_k$ for $k=1,\dots,p$.

Then a standard ADMM algorithm solves the optimization problem \eqref{eq:obj2} by
minimizing the augmented Lagrangian with respect to different variables alternatively.
More explicitly,
at the $(t+1)$-th iteration,
the following updates are implemented:
\begin{subequations}
	\label{eq:upd}
	\begin{align}
	\bm B^{(t+1)}&=\argmin_{\bm B} \left\{\tilde{\ell}_{\mathrm{square}}(\bm B)+\frac{\eta}{2} \|\bm B_\squareun -\bm  D_{0,\squareun}^{(t)} + \bm V_{0,\squareun}^{(t)}\|_F^2 + \frac{\eta}{2}\sum_{k=1}^{p}\left\|{\bm B}_{(k)}-\bm D_{k,(k)}^{(t)} + \bm V_{k,(k)}^{(t)} \right\|_F^2  \right\}, \label{eq:upd1}\\
	\bm D_0^{(t+1)}&=\argmin_{\bm D_0}\left\{\lambda \beta h(\bm D_{0,\squareun})+\frac{\eta}{2} \left\|\bm B^{(t+1)}_\squareun-\bm D_{0,\squareun} + \bm V_{0,\squareun}^{(t)} \right\|_F^2\right\},  \label{eq:upd2}\\
	\bm D_k^{(t+1)}&=\argmin_{\bm D_k}\left\{\lambda \frac{1-\beta}{p}\| \bm D_{k,(k)}\|_*+\frac{\eta}{2} \left\|\bm B^{(t+1)}_{(k)} -\bm D_{k,(k)} + \bm V_{k,(k)}^{(t)} \right\|_F^2\right\} ,\ k=1,\dots,p,\label{eq:upd3}\\
	\bm V_k^{(t+1)}&=\bm V_k^{(t)}+\bm B^{(t+1)}-\bm D_k^{(t+1)},\ k=0,\dots,p,
	\end{align}
\end{subequations}
where $\bm V_k \in \mathbb{R}^{q_1\times\cdots q_{2p}}$, for $k=0,\dots,p$, are scaled Lagrangian multipliers
and $\eta>0$ is an algorithmic parameter.
An adaptive strategy to tune $\eta$ is provided in
\citet{Boyd-Parikh-Chu10}.
One can see that Steps \eqref{eq:upd1}, \eqref{eq:upd2} and \eqref{eq:upd3}
involve additional optimizations.
Now we discuss how to solve them.

The objective function of \eqref{eq:upd1} is a quadratic function, and so we can easily solve this
with a closed-form solution, given in line 2 of Algorithm \ref{admm}.
To solve \eqref{eq:upd2} and \eqref{eq:upd3}, we use proximal operator $\mathrm{prox}^k_v$, $k=1,\dots,p$ and $\mathrm{prox}_v^+ : \mathbb{R}^{q_1\times \cdots\times q_{2p}} \rightarrow \mathbb{R}^{q_1\times \cdots\times q_{2p}} $ respectively defined by
\begin{subequations}
	\begin{align}
	\mathrm{prox}_v^k (\bm A) &= \argmin_{\bm W\in \mathbb{R}^{q_1\times \cdots\times q_{2p}} } \left\{  \frac{1}{2}  \|\bm W_{(k)} - \bm A_{(k)}\|_F^2 + v \|\bm W_{(k)} \|_*\right\}, \label{eq:prox1}\\
	\mathrm{prox}^+_v (\bm A) &= \argmin_{\bm W\in \mathbb{R}^{q_1\times \cdots\times q_{2p}} } \left\{  \frac{1}{2}  \|\bm W_\squareun  - \bm A_\squareun\|_F^2 + v h(\bm W_\squareun)\right\}, \label{eq:prox2}
	\end{align}
	\label{eq:prox}
\end{subequations}
for $v\ge 0$.
By Lemma 1 in \cite{mazumder2010spectral}, the solutions to  \eqref{eq:prox} have closed forms.

For \eqref{eq:prox1},
write the singular value decomposition of
$\bm A_{(k)}$
as
$\bm U\mathrm{diag}((\tilde{a}_1,\dots,\tilde{a}_{q_k})) \bm  V^\tp $, then
$[\mathrm{prox}_v^k (\bm A)]_{(k)} = \bm U \mathrm{diag}(\bm \tilde{\bm c}) \bm V^\tp$
where  $\tilde{\bm c} = (( \tilde{ a}_1-v)_+, (\tilde{ a}_2-v)_+,\dots, (\tilde{ a}_{q_k}-v)_+)$.
As for \eqref{eq:prox2},
is restricted to be a symmetric matrix since the penalty $h$ equals infinity otherwise. Thus \eqref{eq:prox2} is equivalent to minimizing
$\left\{  (1/2)  \|\bm W_\squareun  - (\bm A_\squareun+ \bm A_\squareun^\tp)/2\|_F^2 + v h(\bm W_\squareun)\right\}$ since $\langle \bm W_\squareun,  (\bm A_\squareun- \bm A_\squareun^\tp)/2 \rangle  = \langle (\bm W_\squareun + \bm W_\squareun^\tp)/2,  (\bm A_\squareun- \bm A_\squareun^\tp)/2 \rangle = 0$ .
Suppose
that $(\bm A_\squareun + \bm A_\squareun^\tp)/2$ yields eigen-decomposition $\bm P\mathrm{diag}((\tilde{a}_1,\dots,\tilde{a}_{q}) \bm  P^\tp $.
Then
$[\mathrm{prox}^+_v (\bm A)]_\squareun = \bm P \mathrm{diag}(\bm \tilde{\bm c}) \bm P^\tp$, where  $\tilde{\bm c} = (( \tilde{ a}_1-v)_+, (\tilde{ a}_2-v)_+,\dots, (\tilde{ a}_{q}-v)_+)$.
Unlike singular values, the eigenvalues may be negative. Hence, as opposed to $\mathrm{prox}^k_v$, this procedure $\mathrm{prox}_v^+$ also removes eigen-components with negative eigenvalues.

The details of computational algorithm are given in Algorithm \ref{admm}, an
accelerated version of ADMM which involves additional steps for a faster algorithmic convergence.

\begin{algorithm}[h]
	\SetAlgoLined
	\SetKwInOut{init}{Initialization}
	\KwIn{$\hat{\bm V}_{k}^{(0)}\in \mathbb{R}^{q_1\times\cdots\times q_{2p}}$, $k= 0,1,\dots,p$, and $\bm B^{(0)}\in \mathbb{R}^{q_1\times\cdots\times q_{2p}}$ such that $\hat{\bm V}_{0,(0)}$ and $\bm B^{(0)}_\squareun$ are symmetric matrices;  $\bm M_k = [\bm M_{1,k}^\tp,\dots,\bm M_{n,k}^\tp]^\tp $, $k=1,\dots,p$; $\bm Z_i = (Z_{ijj'})_{1\leq j,j'\leq m}$ , $i=1,\dots, n$;    $\tilde{\bm I} = [I (i\neq j)]_{1\leq i, j \leq m}$;  $\eta > 0$; $T$
	}
	\BlankLine
	\init{ $\alpha_k^{(0)}\leftarrow 1$, $\bm D_k^{(-1)} \leftarrow \bm B^{(0)}$, $\hat{\bm D}_k^{(0)} \leftarrow \bm B^{(0)}$, $\bm V_{k}^{(-1)} \leftarrow \hat{\bm V}_{k}^{(0)}$, $k=0,1,\dots,p$

		$\bm L_i \leftarrow [\bm M_{i,1}^\tp \odot \bm M_{i,2}^\tp\odot\cdots\odot \bm M_{i,p}^\tp]^\tp$, $i=1,\dots,n$, where $\odot$ is the Khatri–Rao product defined as $\bm A \odot \bm B = [a_i \otimes b_i]_{i=1,...,r} \in \mathbb{R}^{r_ar_b \times r}$  for $\bm A \in \mathbb{R}^{r_a \times r}$, $\bm B \in \mathbb{R}^{r_b \times r}$ and $a_i, b_i$ are $i$-th column of matrices $\bm A$ and $\bm B$ respectively.

		$\bm G \leftarrow \frac{1}{nm(m-1)} \sum_{i=1}^n (\bm {L}_i \otimes \bm {L}_i)^\tp   \mathrm{diag}( \mathrm{vec} (\tilde{\bm I})) (\bm {L}_i \otimes \bm {L}_i)$
		$\bm h \leftarrow  \frac{2}{nm(m-1)} \sum_{i=1}^n (\bm {L}_i \otimes \bm {L}_i)^\tp  \mathrm{diag}( \mathrm{vec} (\tilde{\bm I})) \mathrm{vec}(\bm Z_i)$

		$\bm Q \leftarrow (2(\bm G +\frac{p+1}{2}*\eta * \bm I))^{-1}$}
	\BlankLine
	\For{$t=0,1,\dots,T$}
	{
		$\vec (\bm B^{(t+1)}_\squareun) \leftarrow \bm Q \{\bm h +  \eta \sum_{k=0}^{p} \vec( [\bm D_k^{(t)} - \hat{\bm V}_k^{(t)}]_\squareun)\}$\\
		\For{$k=0,1,\dots,p$}{
			\eIf{$k=0$}{
				$\bm D_0^{(t)} \leftarrow \mathrm{prox}^+_{\lambda\beta/ \eta}(\bm B^{(t+1)} + \hat{\bm V}_0^{(t)})$
			}{
				$\bm D_k^{(t)} \leftarrow \mathrm{prox}^k_{\lambda(1-\beta)/(p\eta)}(\bm B^{(t+1)}+\hat{\bm V}_{k}^{(t)})$
			}
			$ \bm V_k^{(t)} \leftarrow  \hat{\bm V}_k^{(t)}+ \bm B^{(t+1)}-\bm D_k^{(t)}$\\
			$\alpha_k^{(t+1)} \leftarrow  \frac{1+\sqrt{1+4(\alpha_k^{(t)})^2}}{2}$\\
			$\hat{\bm D}_k^{(t+1)}  \leftarrow  \bm D_k^{(t)} + \frac{\alpha_k^{(t)}-1}{\alpha_k^{(t+1)}}(\bm D_k^{(k)}-\bm D_k^{(k-1)})$\\
			$\hat{\bm V}_k^{(t+1)} \leftarrow  \bm V_k^{(t)}+\frac{\alpha_k^{(t)}-1}{\alpha_k^{(t+1)}}(\bm V_k^{(t)}-\bm V_k^{(t-1)})$
		}
		Stop if objective value change less than tolerance.
	}
	\KwOut{$\bm D_0^{(T)}$}
	\caption{Accelerated ADMM for solving (\ref{obj})}
	\label{admm}
\end{algorithm}

\section{Asymptotic Properties}
\label{sec:theory}

In this section, we conduct an asymptotic analysis for the proposed estimator $\hat \C$  as defined in \eqref{propmin}.
Our analysis has a unified flavor
such that the derived convergence rate of the proposed estimator
automatically adapts to sparse and dense settings.
Throughout this section,
we neglect the mean function estimation error
by setting $\mu_0(\bm{t}) = \hat{\mu}(\bm{t})= 0$ for any $\bm{t} \in \mathcal{T}$,
which leads to a cleaner and more focused analysis.
The additional error from the mean function estimation can be incorporated
into our proofs without any fundamental difficulty.

\subsection{Assumptions} \label{sec:assum}
Without loss of generality let $\mathcal{T}=[0,1]^p$.
The assumptions needed in the asymptotic results are listed as follows.

\begin{assumption}
	\label{assum_space}
	Sample fields $\{X_i: i=1, \ldots, n\}$ reside in $\Hscr=\bigotimes_{k=1}^p\mathcal{H}_k$
	where $\Hscr_k$ is an RKHS of functions on $[0,1]$ with a continuous and square integrable reproducing kernel $K_k$.
\end{assumption}

\begin{assumption}
	\label{am1}
	The true (folded) covariance function $\C_0 \neq 0$ and $\C_0 \in \Gscr = \bigotimes^d_{j=1} \mathcal{G}_j$, where $d=2p$, $\mathcal{G}_j=\mathcal{H}_j$ for $j=1,\dots,p$ and
	$\mathcal{G}_j =\mathcal{H}_{j-p}$ for $j=p+1,\dots,d$.
\end{assumption}

\begin{assumption}
	\label{assum_loc}
	The locations
	$\{\bm{T}_{ij}:i=1,\dots,n; j=1,\dots, m\}$ are independent
	random vectors from  $\mathrm{Uniform}[0,1]^{p}$,
	and they are independent of
	$\left\lbrace X_i:i=1,\dots,n \right\rbrace$.  \\The errors $\left\lbrace \epsilon_{ij} : i=1,\dots,n ; j=1,\dots,m\right\rbrace$ are independent of both locations and sample fields.
\end{assumption}

\begin{assumption}
	\label{assum_sam}
	For each $\bm t \in \mathcal{T}$, $X(\bm t)$ is sub-Gaussian with
	a parameter $b_X >0$ which does not depend on $\bm t$, i.e., $\E[\exp\left\lbrace \beta X(\bm t)\right\rbrace ]\leq \exp\left\lbrace b_X^2 \beta^2 /2  \right\rbrace $ for all $\beta$ and $\bm t \in \mathcal{T}$.
\end{assumption}

\begin{assumption}
	\label{assum_eps}
	For each $i$ and $j$, $\epsilon_{ij}$ is a mean-zero sub-Gaussian random variable with a parameter $b_\epsilon$ independent of $i$ and $j$, i.e., $\E[\exp\left\lbrace \beta \epsilon_{ij}\right\rbrace ]\leq \exp\left\lbrace b_\epsilon^2 \beta^2 /2  \right\rbrace$.\\
	Moreover
	all errors $\left\lbrace \epsilon_{ij} : i=1,\dots,n ; j=1,\dots,m\right\rbrace$ are independent.
\end{assumption}

Assumption \ref{assum_space} delineates a tensor product RKHS modeling,
commonly seen in the nonparametric regression literature \citep[e.g.,][]{Wahba90, Gu13}.
In Assumption \ref{am1}, the condition $\C_0\in\Gscr$ is satisfied
if
$\E\|X\|_{\Hscr}^2 < \infty$, as shown in \cite{Cai-Yuan10}.
Assumption \ref{assum_loc} is specified for random design and we adopt the uniform distribution here  for simplicity.
The uniform distribution on $[0,1]^p$ can be generalized to any other continuous distribution of which density function $\pi$
satisfies  $c_{\pi} \le \pi(\bm{t})\le c_{\pi}'$
for all $\bm{t}\in [0,1]^p$, for some constants $0<c_{\pi}\le c_{\pi}' < 1$,
to ensure both Theorems \ref{thm_n,m_norm} and \ref{thm_L2} still hold.
Assumptions \ref{assum_sam} and \ref{assum_eps} involve sub-Gaussian conditions of the stochastic process and measurement error,
which are standard tail conditions.

\subsection{Reproducing kernels} \label{sec:reproducing}

In Assumption \ref{assum_space},
the ``smoothness" of the function in the underlying RKHS
is not explicitly specified.
It is well-known that such smoothness conditions are directly
related to the eigen-decay of the respective reproducing kernel.
By Mercer's Theorem \citep{mercer1909xvi},
the reproducing kernel $K_\Hscr((t_1,\ldots,t_{p}),(t'_1,\ldots,t'_{p}))$
of $\Hscr$ possesses the eigen-decomposition
\begin{align}
\label{eqn:kernel}
K_\Hscr((t_1,\ldots,t_{p}),(t'_1,\ldots,t'_{p}))= \sum_{l=1}^{\infty} \mu_l \phi_l (t_1,\ldots,t_{p}) \phi_l (t'_1,\ldots,t'_{p}),
\end{align}
where $\{\mu_l: l \geq 1\}$ are non-negative eigenvalues  and $\{\phi_l: l \geq 1\}$ are $L^2$ eigenfunctions on $[0,1]^p$.
Then for the space $\Hscr \otimes \Hscr$, which is also identified by $\Gscr = \bigotimes_{k=1}^{d} \Gscr_k$,
its corresponding reproducing kernel $K_{\Gscr}$ has the following eigen-decomposition
\begin{align*}
&\quad K_\Gscr((x_1,\ldots,x_{2p}),(x'_1,\ldots,x'_{2p}))\\
&= K_\Hscr((x_1,\ldots,x_{p}),(x'_1,\ldots,x'_{p})) K_\Hscr((x_{p+1},\ldots,x_{2p}),(x'_{p+1},\ldots,x'_{2p}))\\
&= \sum_{l,h=1}^{\infty}  \mu_l \mu_h  \phi_l (x_1,\ldots,x_{p}) \phi_h (x_{p+1},\ldots,x_{2p})   \phi_l (x'_1,\ldots,x'_{p}) \phi_h (x'_{p+1},\ldots,x'_{2p}),
\end{align*}
where $\{\mu_l\mu_h: l,h \geq 1\}$ are the eigenvalues of $K_\Gscr$.
Due to continuity assumption (Assumption \ref{assum_space}) of the univariate kernels,
there exists a constant $\kc$ such that

$$
\sup_{(x_1,\ldots,x_{2p})\in [0,1]^{2p}}K_\Gscr ((x_1,\ldots,x_{2p}),(x_1,\dots,x_{2p}))
\leq \kc.
$$

The decay rate of the eigenvalues $\{\mu_l\mu_h: l, h \geq 1\}$ is involved in our analysis through two quantities $\kappa_{n,m}$ and $\eta_{n,m}$,
which have relatively
complex forms
defined in Appendix \ref{app:kappa_and_eta}.
Similar quantities can be found in other analyses of RKHS-based estimators
\citep[e.g.,][]{Raskutti-Wainwright-Yu12}
that accommodate general choices of RKHS.
Generally
$\kappa_{n,m}$ and $\eta_{n,m}$ are expected to diminish
in certain orders of $n$ and $m$,
characterized by the decay rate of the eigenvalues $\{\mu_l\mu_h\}$.
The smoother the functions in the RKHS,
the faster these two quantities diminish.
Our general results in Theorems \ref{thm_n,m_norm} and \ref{thm_L2}
are specified in terms of these quantities.
To provide a solid example,
we derive the orders of $\kappa_{n,m}$ and $\eta_{n,m}$ under a Sobolev-Hilbert space
setting
and provide the convergence rate of the proposed estimator in Corollary \ref{cor_sob}.

\subsection{Unified rates of convergence} \label{sec:rate}

We write the penalty in \eqref{propmin} as
$I(\C) = \beta \|\C_\squareun\|_* +(1-\beta)p^{-1} \sum_{k=1}^p \|\C_{(k)}\|_*$.
For arbitrary functions $g_1,g_2\in \Gscr$,  define their empirical inner product and the corresponding (squared) empirical norm as
$$
\langle g_1, g_2 \rangle_{n,m}= \frac{1}{nm(m-1)}\sum_{i=1}^{n}\sum_{1\leq j,j' \leq m} g_1(T_{ij1}, \dots, T_{ijp},T_{ij'1}, \dots, T_{ij'p}) g_2(T_{ij1}, \dots, T_{ijp},T_{ij'1}, \dots, T_{ij'p}),$$ $$\|g_1\|_{n,m}^2= \langle g_1,g_1\rangle_{n,m}.$$
Additionally, the $L^2$ norm of a function $g$ is defined as
$\|g\|_2=\{\int_{\mathcal{T}} g^2(\bm t)\, d \bm t \} ^{1/2}$.

Define
$\xi_{n,m} = \max\lbrace \eta_{n,m}, \kappa_{n,m}, ( n^{-1} \log n )^{1/2}\rbrace$.
We first provide the empirical $L^2$ rate of convergence for $\hat \C$.

\begin{thm}
	\label{thm_n,m_norm}
	Suppose that Assumptions \ref{assum_space}--\ref{assum_eps} hold.
	Assume $\xi_{n,m}$ satisfies $(\log n)/n \leq \xi_{n,m}^2/(\log\log \xi_{n,m}^{-1})$.
	If $\lambda \ge L_1 \xi_{n,m}^2$ for some constant
	$L_1>0$ depending on $b_X$, $b_\epsilon$ and $\kc$,
	we have
	\begin{align*}
	\|\hat \C - \C_0\|_{n,m} \leq  \sqrt{ 2 I(\C_0)\lambda}+ L_1 \xi_{n,m},
	\end{align*}
	with probability at least $1-\exp(-\uni n\xi_{n,m}^2/ {\log n} )$ for some positive universal constant $\uni$.
\end{thm}

Next, we provide the $L^2$ rate of convergence for $\hat \C$.
\begin{thm}
	\label{thm_L2}
	Under the same conditions as Theorem \ref{thm_n,m_norm},  there exist a positive constant $L_2$ depending on $b_X$, $b_\epsilon$, $b$ and $I(\C_0)$, such that
	$$\|\hat \C - \C_0\|_2 \leq 2\sqrt{ I(\C_0)\lambda} + L_2 \xi_{n,m},$$
	with probability at least $1-\exp(-\pc n\xi_{n,m}^2/\log n)$ for some constant $\pc$ depending on $b$.
\end{thm}

\supp{The proofs of Theorems \ref{thm_n,m_norm} and \ref{thm_L2} can be found in Section S1 in the supplementary material.}

\begin{remark}\label{rmk:theory.1}
	Theorems \ref{thm_n,m_norm} and \ref{thm_L2}
	are applicable to general RKHS $\mathcal{H}$ which satisfies Assumption \ref{assum_space}.
	The convergence rate depends on the eigen-decay rates of the reproducing kernel.
	A special case of polynomial decay rates for univariate RKHS will be given in Corollary \ref{cor_sob}.
	Moreover, our analysis has a \textit{unified} flavor in the sense that the resulting convergence
	rates automatically adapt to the orders of both $n$ and $m$.
	In Remark \ref{rmk:regime.1} we will provide a discussion of a ``phase transition" between dense and sparse functional data revealed by our theory.
\end{remark}
\begin{remark}\label{rmk:theory.2}
	With a properly chosen $\lambda$,
	Theorems \ref{thm_n,m_norm} and \ref{thm_L2}
	bound the convergence rates (in terms of both the empirical and theoretical $L^2$ norm)
	by $\xi_{n,m}$, which cannot be faster than $(n^{-1}\log n)^{1/2}$.
	The logarithmic order is due to the use of Adamczak bound in \supp{Lemma
	S2 in the supplementary material.}
	If one further assumes boundedness for the sample fields $X_i$'s (in terms of the sup-norm) and the noise variables $\epsilon_{ij}$'s,
	we can instead use
	Talagrand concentration inequality (Bousquet bound in \cite{koltchinskii2011oracle}) and the results in 
	Theorems \ref{thm_n,m_norm} and \ref{thm_L2} can be improved to
	$\max\{\|\hat \C - \C_0\|^2_{n,m} , \|\hat \C - \C_0\|^2_{2}\} = \bigOp( \tilde{\xi}_{n,m}^{2})$,
	where $\tilde{\xi}_{n,m} = \max \{ \eta_{n,m}, \kappa_{n,m}  , n^{-1/2}\}$.

\end{remark}

Next we focus on a special case
where the reproducing kernels of the univariate RKHS $\Hscr_k$'s
exhibit polynomial eigen-decay rates,
which holds for a range of commonly used RKHS.
A canonical example is $\alpha$-th order Sobolev-Hilbert space:
\begin{equation*}
\Hscr_k = \{f: f^{(r)}, r = 0,\dots,\alpha, \mathrm{are\  absolutely\  continuous}; f^{(\alpha)}\in L^2([0,1])\},
\end{equation*}
where $k=1,\dots,p$.  Here $\alpha$ is the same as $\alpha$ in Corollary \ref{cor_sob}.
To derive the convergence rates,
we relate the eigenvalues $\nu_l$ in \eqref{eqn:kernel} to the univariate RKHS $\Hscr_k$, $k=1,\dots,p$.
Due to Mercer's Theorem, the reproducing kernel $K_k$ of $\Hscr_k$ yields an eigen-decomposition with non-negative eigenvalues $\{\mu_l^{(k)}: l\geq 1\}$ and an $L^2$ eigenfunction $\{ \phi^{(k)}_l: l \geq 1 \}$, i.e.,
$
K_k(t,t') = \sum_{l=1}^\infty \mu_l^{(k)} \phi_l^{(k)} (t) \phi^{(k)}_l(t').
$
Therefore, the set of eigenvalues $\{\mu_l:l \ge 1\}$ in \eqref{eqn:kernel}
is the same as the set
$\{\prod_{k=1}^{p} \mu_{l_k}^{(k)}: l_1,\dots, l_p\ge 1\}$.
Given the eigen-decay of $\mu_l^{(k)}$,
one can obtain the order of $\xi_{n,m}$ and hence the convergence rates from Theorems \ref{thm_n,m_norm} and \ref{thm_L2}.
Here are the results under the setting of a polynomial eigen-decay.

\begin{corollary}
	\label{cor_sob}
	Suppose that the same conditions in Theorem \ref{thm_L2} hold. If the eigenvalues of $K_k$ for $\Hscr_k, k=1, \ldots, p,$ have polynomial decaying rates, that is, there exists $\alpha>1/2$ such that $\mu_l^{(k)} \asymp  l^{-2\alpha }$ for all $k=1,\dots,p$, then
	$$
	\max\left\{\|\hat \C - \C_0\|^2_{n,m},  \|\hat \C - \C_0\|^2_{2}\right\}
	= \bigOp\left( \max\left\{ (nm)^{-\frac{2\alpha }{1+2\alpha }}\{\log( nm)\}^{\frac{2\alpha(2p-1)}{2\alpha +1 }}, \frac{\log n}{n}\right\}\right).
	$$
\end{corollary}

\begin{remark}\label{rmk:regime.1}
	All Theorems \ref{thm_n,m_norm} and \ref{thm_L2} and Corollary \ref{cor_sob} reveal a ``phase-transition'' of the convergence rate depending on the relative magnitudes between $n$, the sample size, and $m$, the number of observations per field.
	When $ \kappa^2_{n,m} \ll  (\log n/n)$, which is equivalent to $m \gg n^{1/(2\alpha )}(\log n )^{2p-2-1/(2\alpha)}$ in Corollary \ref{cor_sob},
	both empirical and theoretical $L^2$ rates of convergence can achieve the near-optimal rate $\sqrt{\log n/n}$.
	Under the stronger assumptions in Remark \ref{rmk:theory.2}, the convergence rate will achieve the optimal order $\sqrt{1/n}$ when $\kappa^2_{n,m} \ll  1/n$ (or $m \gg n^{1/(2\alpha )}(\log n )^{2p-1}$ in Corollary \ref{cor_sob}). In this case, the observations are so densely sampled that we can estimate the covariance function as precisely as if the entire sample fields are observable.
	On the contrary, when  $\kappa^2_{n,m} \gg  (\log n/n)$ (or $m \ll n^{1/(2\alpha )}(\log n )^{2p-2-1/(2\alpha)}$ in Corollary \ref{cor_sob}),
	the convergence rate is determined by the total number of observations $nm$.
	When $p=1$, the asymptotic result
	in Corollary \ref{cor_sob}, up to some $\log m$ and $\log n$ terms, is the same as the minimax optimal rate
	obtained by \citet{Cai-Yuan10},
	and is comparable to the $L^2$ rate obtained by \citet{paul2009consistency} for $\alpha=2$.
\end{remark}

\begin{remark}\label{rmk:regime.2}
	For covariance function estimation for unidimensional functional data, i.e., $p=1$, a limited number of approaches, including \cite{Cai-Yuan10}, \cite{li2010uniform}, \cite{zhang2016sparse}, and \cite{liebl2019inference},
	can achieve unified theoretical results in the sense that they hold for all relative magnitudes of $n$ and $m$. The similarity of these approaches is the availability of a closed form for each covariance function estimator.
	In contrast, our estimator obtained from \eqref{propmin} does not have a closed form due to the non-differentiability of the penalty, but it can still achieve unified theoretical results which hold for both unidimensional and multidimensional functional data. Due to the lack of a closed form of our covariance estimator, we used the empirical process techniques \citep[e.g.,][]{bartlett2005local, koltchinskii2011oracle} in the theoretical development.
	In particular, we have developed a novel grouping lemma \supp{(Lemma
	S4 in the supplementary material)} to \textit{deterministically} decouple the dependence within a $U$-statistics of order 2.
	We believe this lemma is of independent interest.
	In our analysis, the corresponding $U$-statistics is \textit{indexed} by a function class,
	and this grouping lemma provides a tool to obtain uniform results \supp{(see Lemma S3 in the supplementary material)}.
	In particular, this allows us to relate the empirical and theoretical $L^2$ norm of the underlying function class, in precise enough order dependence on $n$ and $m$ to derive the unified theory.
	See Lemma S3 for more details.
	To the best of our knowledge, this paper is one of the first in the FDA literature that derives a unified result in terms of empirical process theories,
	and the proof technique is potentially useful for some other estimators without a closed form.
\end{remark}

  \section{Simulation}
\label{sec:simulation}
To evaluate the practical performance of the proposed method, we conducted a simulation study. We in particular focused on two-dimensional functional data.
Let $\mathcal{H}_1$ and $\mathcal{H}_2$ both
be the RKHS with kernel $K(t_1,t_2)=\sum_{k =1}^\infty (k \pi)^{-4}e_k(t_1)e_k(t_2)$,
where $e_k(t) = \sqrt{2} \cos (k\pi t)$, $k \geq 1$.
This RKHS
has been used in various studies in FDA, e.g., the simulation study of \citet{Cai-Yuan12}.
Each $X_i$
is generated from
a mean-zero Gaussian random field
with
a covariance function
$$
\Cs_{0}((s_1,s_2), (t_1,t_2))=\C_0(s_1,s_2,t_1,t_2)=\sum^R_{k=1} k^{-2} \psi_k(s_1,s_2)\psi_k(t_1,t_2),
$$
where the eigenfunctions $\psi_k(t_1,t_2) \in
\mathcal{P}_{r_1,r_2}:=\{e_i(t_1)e_j(t_2): i=1,\dots, r_1; j=1,\dots, r_2\}$.
Three
combinations of one-way ranks ($r_1, r_2$) and two-way rank $R$ were studied for $\C_0$:
\begin{center}
	\textbf{Setting 1}: $R=6$, $r_1=3$, $r_2=2$; \hspace{0.5cm} \textbf{Setting 2}: $R=6$, $r_1=r_2=4$;\\
	\textbf{Setting 3}: $R=r_1=r_2=4$.
\end{center}
For each setting, we chose $R$ functions out of $\mathcal{P}_{r_1,r_2}$ to be
$\{\psi_k\}$ such that smoother functions are associated with larger eigenvalues. The details
are described in \supp{Section S2.1 of the supplementary material.}

In terms of sampling plans, we considered both sparse and dense designs.
Due to the space limit, here we only show and discuss the results for the
sparse design, while defer those for the dense design  to \supp{Section S2.3 of the supplementary material.}
For the sparse design, the random locations $\bm T_{ij}, j=1, \ldots, m,$ were independently generated from the continuous uniform distribution on $[0,1]^2$ within each field and across different fields, and the random errors $\{\epsilon_{ij}: i=1, \ldots, n; j=1, \ldots, m\}$ were independently generated from $N(0,\sigma^2)$.
In each of the 200 simulation runs, the observed data were obtained following \eqref{eqn:obsmodel} with various combinations of $m=10,20$,
$n=100,200$ and noise level $\sigma = 0.1, 0.4$.

We compared the proposed method, denoted by \convex, with three existing
methods:   1) $\Wong$: the estimator based on \cite{Wong-Zhang19} with
adaption to two dimensional case (see Section \ref{sec:setup});
2) \llsm:  local linear smoothing with Epanechnikov kernel;
3)  \llsmp: the two-step estimator constructed by retaining eigen-components
of
\llsm   selected by 99\% fraction of variation explained (FVE),
and then removing the eigen-components with negative eigenvalues.
For both $\Wong$ and $\convex$,
5-fold cross-validation was adopted to select the corresponding tuning parameters.

Table \ref{ran_n200} show the average integrated squared error (AISE), average
of estimated two-way rank ($\bar R$), as well as average of estimated one-way
ranks ($\bar r_1, \bar r_2$)  of the above covariance estimators over 200
simulated data sets in respective settings when sample size is $n=200$.
Corresponding results for $n=100$  can be found in  \supp{Table
	S4 of the supplementary material}, and they lead to similar
conclusions.
Obviously  \llsm{} and \llsmp{}, especially \llsm,  perform significantly worse than the other two methods in both estimation accuracy and rank reduction (if applicable). Below we only compare $\convex$ and $\Wong$.

Regarding estimation accuracy, the proposed $\convex$ has uniformly smaller AISE values than $\Wong$,
with around $10\% \sim 20\%$ improvement of AISE over $\Wong$
in most cases under Settings 1 and 2.
If the standard
error (SE) of AISE is taken into account,
the improvements of AISE by $\convex$
are more distinguishable in Settings 1 and 2 than those in Setting 3
since the SEs of $\Wong$ in Setting 3 are relatively high. This is due to the fact that
in Setting 3, marginal basis  are not shared by different two-dimensional eigenfunctions, and
hence $\convex$ cannot benefit from the structure sharing among eigenfunctions. Setting 3 is in fact an extreme setting we designed to challenge the proposed method.

For rank
estimation, $\Wong$ almost always underestimates two-way ranks,
while $\convex$ typically overestimates both one-way and two-way ranks.
For $\convex$, the average one-way rank estimates are always smaller than the average two-way rank estimates, and their differences are substantial in Settings 1 and 2. This demonstrates the benefit of $\convex$ of detecting structure sharing of one-dimensional basis among two-dimensional eigenfunctions.

We also tested the performance of $\convex$ in the dense and regular designs, and compared it with the existing methods mentioned above together with
the one by \cite{wang2017regularized},
which is not applicable to the sparse design. Details are given in \supp{Section S2.3 of the supplementary material}, where
all methods achieve similar AISE values,
but $\convex$ performs slightly better in estimation accuracy when the noise level is high.

\begin{table}[ht]
	\centering
	\caption{Simulation results for three Settings with the sparse design when sample size ($n$) is 200. The AISE values with standard errors (SE) in parentheses are provided for the four covariance estimators in comparison, together with average two-way ranks ($\bar R$) for those estimators which can lead to rank reduction (i.e., \convex, \Wong, and \llsmp) and average one-way ranks ($r_1$, $r_2$) for $\convex$. }
	\label{ran_n200}
	\resizebox{\textwidth}{!}{%
		{\small
			\begin{tabular}{crr | l llll}
				\hline
				Setting & $m$ & $\sigma$ &  & $\convex$ & $\Wong$ & \llsm  & \llsmp \\
				\hline
				1 & 10 & 0.1 & AISE & 0.053 (1.97e-03) & 0.0632 (3.22e-03) & 0.652 (1.92e-01) & 0.337 (5.35e-02)\\
				&  &  & $\bar R$& 8.38 & 2.94 & - &172.70 \\
				&  &  &$\bar r_1$, $\bar r_2$  & 5.4, 5.4& \_ & \_&\_ \\
				\hline
				&  & 0.4 & AISE & 0.0547 (2.01e-03) & 0.0656 (2.72e-03) & 0.714 (2.11e-01) &0.366 (5.96e-02)\\
				&  &  & $\bar R$& 9.16 & 2.84  & - &177.3 \\
				&  &  & $\bar r_1$, $\bar r_2$  & 5.34, 5.32 & \_ & \_ &\_\\
				\hline
				& 20 & 0.1 & AISE & 0.0343 (1.46e-03) & 0.0421 (1.97e-03) & 0.297 (1.39e-02) & 0.206 (4.62e-03)\\
				&  &  & $\bar R$ & 8.38 & 3.78 & - &317.44\\
				&  &  & $\bar r_1$, $\bar r_2$  & 5.84, 5.82 & \_ & \_ &\_\\
				\hline
				&  & 0.4 & AISE & 0.0354 (1.52e-03) & 0.044 (2.21e-03) & 0.325 (1.58e-02) & 0.223 (4.94e-03)\\
				&  &  & $\bar R$ & 8.86 & 3.76 & - & 326.31\\
				&  &  & $\bar r_1$, $\bar r_2$  & 5.83, 5.84& \_ & \_ &\_\\
				\hline
				2 & 10 & 0.1 & AISE & 0.0532 (1.98e-03) & 0.0636 (3.12e-03) & 2.33 (1.13e+00) &0.795 (2.98e-01)\\
				&  &  & $\bar R$ & 8.48 & 3.02 & - &191.175 \\
				&  &  & $\bar r_1$, $\bar r_2$& 5.82, 5.82 & \_ & \_ &\_\\
				\hline
				&  & 0.4 & AISE & 0.0548 (2.05e-03) & 0.0686 (3.53e-03) & 2.44 (1.17e+00) & 0.828 (3.04e-01)\\
				&  &  & $\bar R$ & 9.04 & 3.04 & - & 196.34\\
				&  &  & $\bar r_1$, $\bar r_2$ & 5.71, 5.74 & \_ & \_ &\_\\
				\hline
				& 20 & 0.1 & AISE & 0.0341 (1.43e-03) & 0.0419 (2.02e-03) & 0.301 (1.58e-02) &0.208 (4.50e-03)\\
				&  &  & $\bar R$ & 8.99 & 3.74 & - & 318.645\\
				&  &  & $\bar r_1$, $\bar r_2$ & 5.93, 5.92 & \_ & \_ &\_\\
				\hline
				&  & 0.4 & AISE & 0.0348 (1.43e-03) & 0.043 (2.22e-03) & 0.328 (1.78e-02) & 0.225 (4.74e-03)\\
				&  &  &$\bar R$& 8.01 & 3.6 & - & 327.395\\
				&  &  &$\bar r_1$, $\bar r_2$& 5.94, 5.93 & \_ & \_ &\_ \\
				\hline
				3 & 10& 0.1 & AISE & 0.058 (2.62e-03) & 0.0692 (5.33e-03) & 0.454 (7.28e-02) & 0.286 (2.89e-02) \\
				&  &  & $\bar R$ & 6.26 & 3.12 & -& 182.74 \\
				&  &  &  $\bar r_1$, $\bar r_2$ & 5, 5.06 & \_ & \_ & \_ \\
				\hline
				&  & 0.4 & AISE & 0.0598 (2.68e-03) & 0.0733 (6.14e-03) & 0.531 (1.07e-01) &0.323 (4.23e-02)\\
				&  &  & $\bar R$ & 6.48 & 3.2 & - & 185.82 \\
				&  &  &  $\bar r_1$, $\bar r_2$ & 4.99, 5.07 & \_ & \_ & \_ \\
				\hline
				& 20& 0.1 & AISE & 0.0422 (1.37e-03) & 0.0535 (2.64e-03) & 0.267 (5.04e-03) & 0.196 (3.59e-03)\\
				&  &  & $\bar R$& 6.29 & 4.49 & - & 332.09 \\
				&  &  &  $\bar r_1$, $\bar r_2$ & 5.62, 5.69 & \_ & \_ & \_ \\
				\hline
				&  & 0.4 & AISE & 0.0424 (1.30e-03) & 0.0494 (2.42e-03) & 0.292 (5.30e-03) &0.212 (3.72e-03)\\
				&  &  & $\bar R$ & 5.68 & 3.36 & - &338.725 \\
				&  &  &  $\bar r_1$, $\bar r_2$ & 5.59, 5.66 & \_ & \_ & \_ \\
				\hline
	\end{tabular}}}
\end{table}

	\section{Real Data Application}
\label{sec:real}
\begin{figure}[h]
	\centering
	\includegraphics[width=1\linewidth,height=0.32\linewidth]{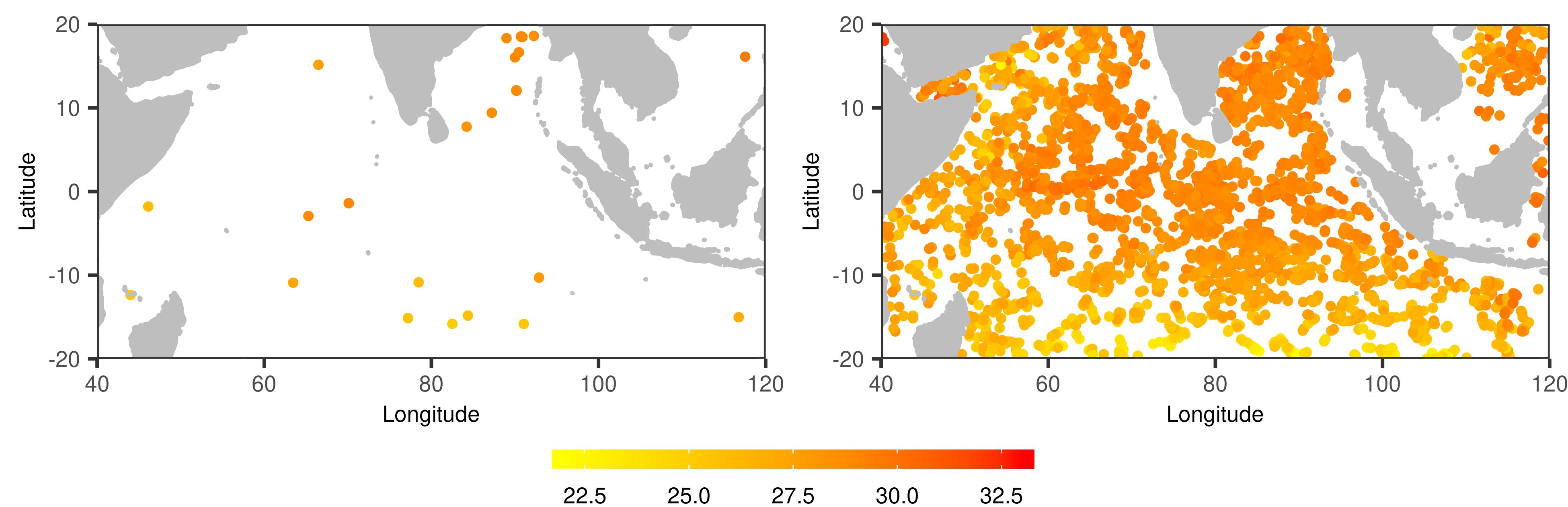}
	\caption{Observations on 2013/09/04 (left), and all observations in
		the data set (right). Points on the map indicate locations (Longitude,
		Latitude) of observations and the color scale of every point shows the
		corresponding Celsius temperature.}
	\label{fig:obs}
\end{figure}

We applied the proposed method to an Argo profile data set,
obtained from \url{http://www.argo.ucsd.edu/Argo_data_and.html}.
The Argo project has a global array of approximately 3,800 free-drifting profiling floats,
which measure temperature and salinity of the ocean.
These floats drift freely in the depths of the ocean most of the time, and ascend regularly to the sea surface,
where they transmit the collected data to the satellites.
Every day only a small subset of floats show up on the sea surface.
Due to the drifting process,
these floats measure temperature and salinity
at irregular locations over the ocean.
See Figure \ref{fig:obs} for examples.

In this analysis, we focus on the different changes of sea surface temperature between the tropical western and eastern Indian Ocean,
which is called the Indian Ocean Dipole (IOD). The IOD is known to be
associated with droughts in  Australia \citep{ummenhofer2009causes} and has a significant effect on rainfall patterns in southeast Australia \citep{behera2003influence}.
According to \cite{shinoda2004surface},  the IOD phenomenon is a predominant inter-annual variation of sea surface temperature  during late boreal summer and autumn \citep{shinoda2004surface},
so in this application
we focused on the sea surface temperature in the Indian Ocean region of longitude 40$\sim$120 and latitude -20$\sim$20
between September and November every year
from 2003 to 2018.

Based on a simple autocorrelation analysis on the gridded data, we decided to use measurements for every ten days in order to reduce the temporal dependence among the data.

At each location of a float on a particular day, the average temperature
between 0 and 5 hPa from the float
is regarded as a measurement.
The Argo float dataset provides multiple versions of data, and we adopted the quality controlled (QC) version.
Eventually we have a two-dimensional functional data collected of $n=144$ days, where the number of observed locations $\bm T_{ij}=(\text{longitudie}, \text{latitude})$ per day
varies from 7 to 47, i.e., $7\leq m_i\leq 47$, $i=1,...,n$, with an average of 21.83.
The locations are rescaled to $[0, 1] \times [0, 1]$. As shown in Figure \ref{fig:obs}, the data
has a random sparse design.

First we used kernel ridge regression
with the corresponding kernel for  the tensor product of two second order
Sobolev spaces \citep[e.g.,][]{Wong-Zhang19}
to obtain a mean function estimate for every month.
Then we applied the proposed covariance function estimator with the same kernel.

\begin{figure}[H]
	\centering
	\includegraphics[width=1\linewidth,height=0.32\linewidth]{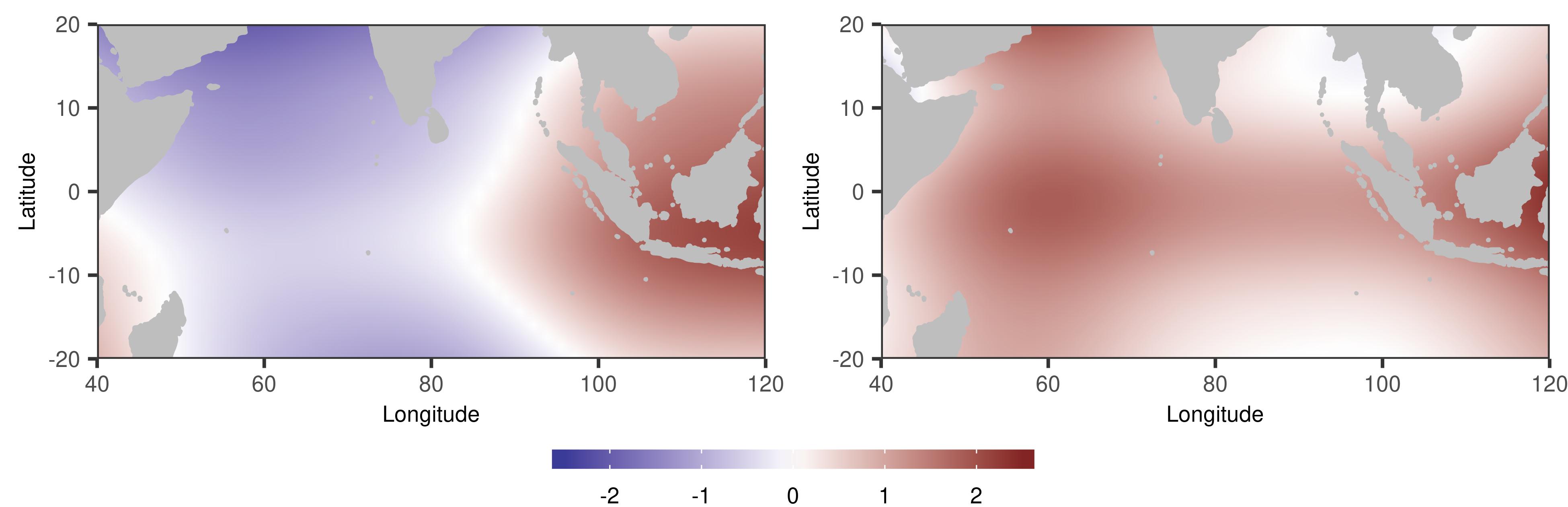}
	\caption{The first two-dimensional $L^2$ eigenfunction (left) and the second  two-dimensional $L^2$ eigenfunction (right). The first eigenfunction explains $33.60\%$ variance and the second eigenfunction explains $25.94\%$ variance. }
	\label{fig:eigfc}
\end{figure}

The estimates of the top two two-dimensional $L^2$ eigenfunctions are illustrated in Figure \ref{fig:eigfc}. The first eigenfunction shows the east-west dipole mode, which aligns with existing scientific findings \citep[e.g.,][]{shinoda2004surface, chu2014future, deser2010sea}.
The second eigenfunction can be interpreted as the basin-wide mode, which is a dominant mode
all around the year \citep[e.g.,][]{deser2010sea, chu2014future}.

To provide a clearer understanding of the covariance function structure, we derived a marginal $L^2$ basis along longitude and latitude respectively. The details are given in Appendix \ref{app:L2_trans}.
The left panel of Figure \ref{fig:onefc} demonstrates that
the first longitudinal marginal basis
reflects a large variation in the western region while the second one corresponds to the variation in the eastern region.
Due to different linear combinations, the variation along longitude may contribute to not only opposite changes between the eastern and western sides of the Indian Ocean as shown in the first two-dimensional eigenfunction, but also an overall warming or cooling tendency as shown in the second two-dimensional eigenfunction. The second longitudinal marginal basis reveals that
the closer to the east boundary, the greater the variation is,
which suggests that the IOD may be related to
the Pacific Ocean. This aligns with the evidence that the IOD has a link with El Ni\tex{\~n}o Southern
Oscillation (ENSO)   \citep{stuecker2017revisiting}, an irregularly
periodic variation in sea surface temperature over the tropical eastern
Pacific Ocean.
As shown in the right panel of Figure \ref{fig:onefc},
the overall trend for the first latitude marginal basis is almost a constant function.
This provides evidence that the IOD is primarily associated with the variation along longitude.

\begin{figure}[H]
	\centering
	\begin{subfigure}[b]{0.49\linewidth}
		\includegraphics[width=\linewidth,height=0.65\linewidth]{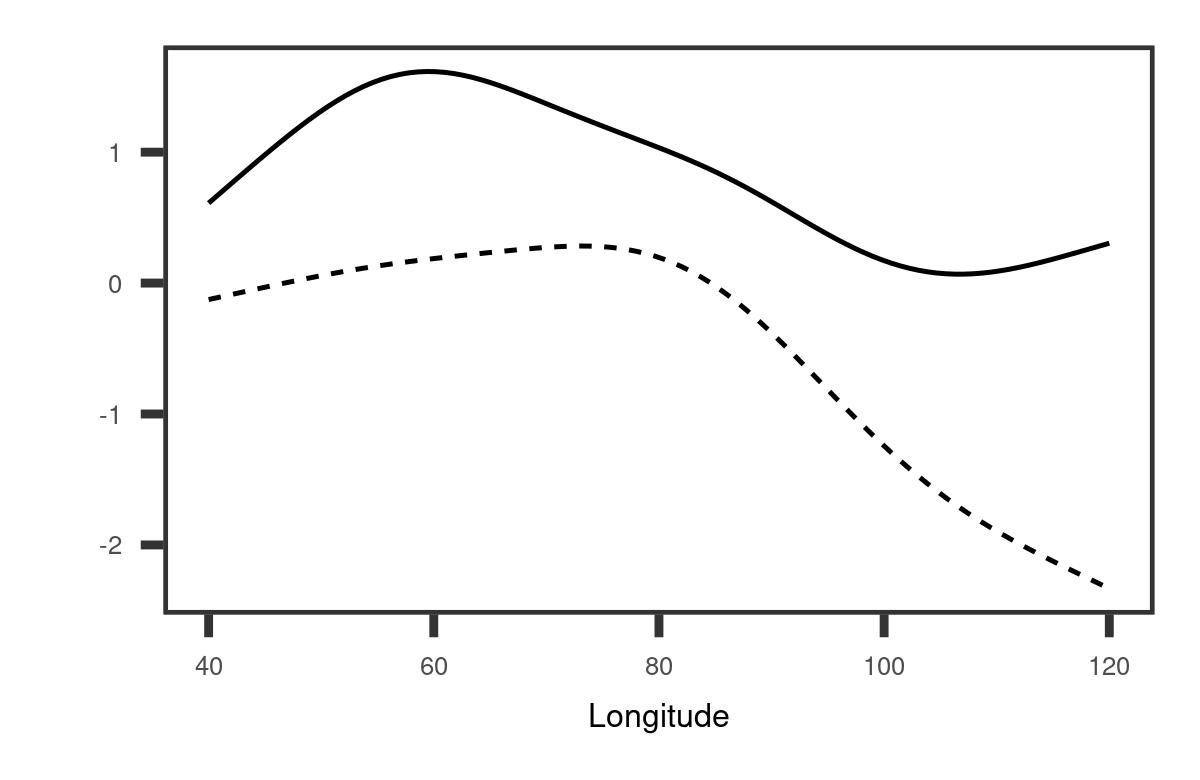}
		\caption{Longitude ($39.06\%$, $36.10\%$)}
	\end{subfigure}
	\begin{subfigure}[b]{0.49\linewidth}
		\includegraphics[width=\linewidth,height=0.65\linewidth]{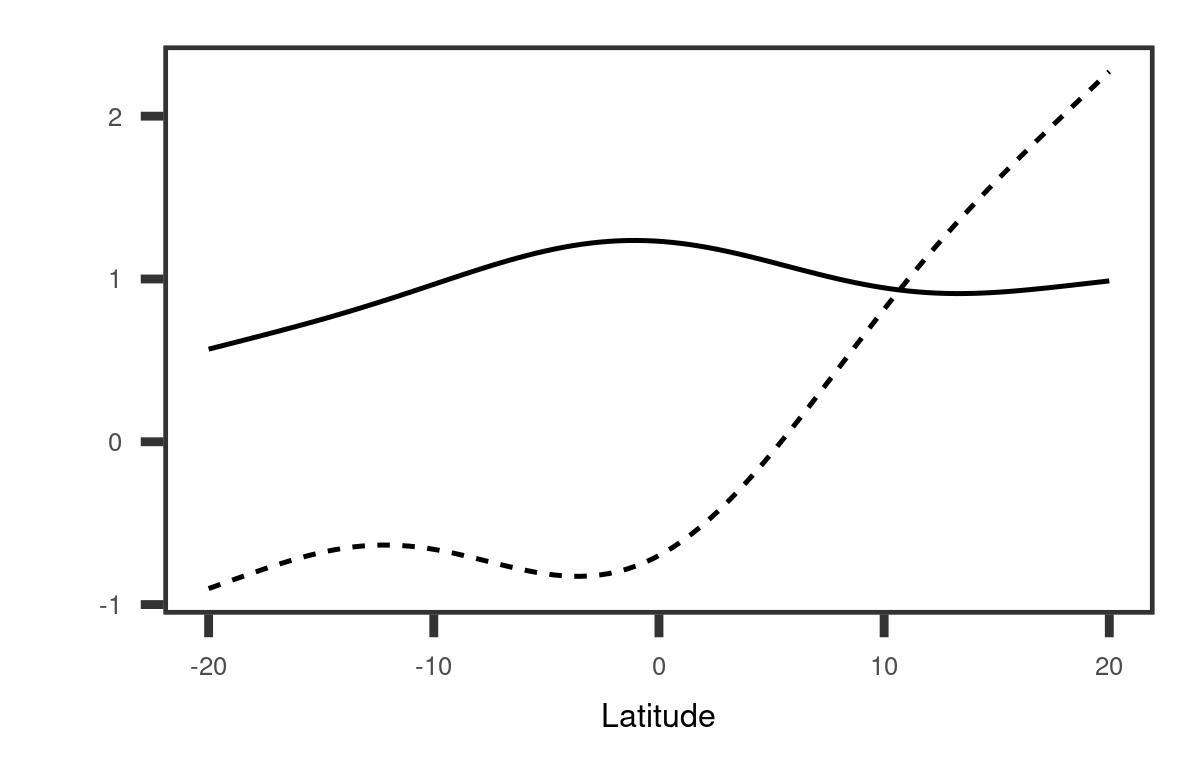}
		\caption{Latitude ($48.22\%$, $25.40\%$)}
	\end{subfigure}
	\caption{The first two marginal $L^2$ basis functions along longitude and latitude respectively. Solid lines are the first marginal basis function and dotted lines are the second marginal basis function. The fractions of variation explained by the corresponding principle components are given in parentheses.
	}
	\label{fig:onefc}
\end{figure}

\clearpage

\appendix

\section*{Appendix}
\section{$L^2$ eigensystem and $L^2$ marginal  basis}
\label{app:L2_trans}
In this section,  we present a transformation procedure to produce $L^2$ eigenfunctions and corresponding eigenvalues from our estimator $\hat {\bm B}$
obtained by \eqref{obj}.

Let $\bm Q_k =[  \int_{[0,1]} K(s,T_{ijk})K(s,T_{i'j'k})ds ]_{1\leq i,i' \leq n, 1\leq j,j' \leq m}$, $k=1, \ldots, p$.
Then $\bm Q_k = \bm M_k \bm R_k \bm M_k^{\tp}$,  where $\bm R_k =[  \int_{[0,1]} v_l(s) v_h (s)ds ] _{1\leq l,h \leq q_k}$ and $\{v_l: l=1, \ldots, q_k\}$ form a basis
of $\Hscr_k$, so
$\bm R_k =\bm M_k^+ \bm Q_k (\bm M_k ^+)^{\tp}$.
The $L^2$ eigenvalues of ${\hat \C}_\squareun$ coincide with the eigenvalues of matrix $\hat{\bm B}^{L}_{\mathrm{square}}:= (\bm R_1 \otimes \ldots \otimes \bm R_p)^{1/2} \hat{\bm B}_\squareun [(\bm R_1 \otimes \ldots \otimes \bm R_p)^{1/2}]^{\tp}$,  and the number of nonzero eigenvalues is the same as the rank of $\hat {\bm B}_\squareun$.  The $L^2$ eigenfunction $\hat \phi_l$ that corresponds to the $l$-th eigenvalue of $\hat \C_\squareun$ can be expressed as
$\hat \phi_l(s_1,...,s_p) = \bm u_l ^{\tp} [ \bm z_1(s_1) \otimes\ldots\otimes \bm z_p(s_p) ] $,
where $\bm z_k (\cdot)$, $k =1,\ldots,p$ are defined in Theorem  \ref{thm_re}, and $\bm u_l = (\bm M_1^+ \otimes\ldots\otimes \bm M_p^+)^{\tp}(\bm R_1\otimes\ldots\otimes\bm R_p)^{-1/2}\bm v_l$ with $ \bm v_l$ being the  $l$-th eigenvector of matrix $\hat{\bm B}^{L}_{\mathrm{square}}$. Using the property of Kronecker products, we have $\hat \phi_l(s_1,...,s_p) = \bm v_l ^{\tp} [ (\bm R_1^{-1/2}\bm M_1^+ \bm z_1(s_1)) \otimes \ldots \otimes (\bm R_p^{-1/2}\bm M_p^+ \bm z_p(s_p)) ]$.

By simple verification, we can see that $\bm R_k^{-1/2} \bm M_k^+ \bm z_k(\cdot)$ are $q_k$ one-dimensional orthonormal $L^2$ functions for dimension $k$, $k=1,...,p$. Therefore, we can also express $\hat \C$ with these $L^2$ one-dimensional basis and the coefficients will form a $2p-$th order tensor of dimension $q_1 \times \ldots q_p \times q_1 \times \ldots q_p$. We use ${\hat{\bm B}}^L$ to represent this new coefficient tensor and extend our unfolding operators to $L^2$ space.
It is easy to see that ${\hat{\bm B}}^L_\squareun = {\hat{\bm B}}^L_{\mathrm{square}}$.

Since $\hat \C_{(k)}$ is a compact operator in the $L^2$ space, this yields a singular value decomposition which leads to a $L^2$ basis characterizing the marginal variation along the $k-$th dimension.
We call it a \emph{$L^2$ marginal  basis} for the $k-$th dimension. Obviously the marginal basis function
$\hat \psi^k_l$ corresponding to the $l$-th singular value for dimension $k$ can be expressed as $\hat \psi^k_l(\cdot) = \bm u^k_l \bm z_k(\cdot)$, where $\bm u^k_l= (\bm M_k^+)^\tp \bm R_k^{-1/2} \bm v^k_l$, and $\bm v^k_l$ is the $l$-th singular vector of $\hat {\bm B}^L_{(k)}$.
And the $L^2$ marginal  singular values of $\hat \C_{(k)}$  coincide with the singular values of matrix $\hat {\bm B}^L_{(k)}$.

\section{Definitions of $\kappa_{n,m}$ and $\eta_{n,m}$}
\label{app:kappa_and_eta}
Here we provide the specific forms of $\kappa_{n,m}$ and $\eta_{n,m}$, which are
closely related to the decay of $\{\mu_l\mu_h: l,h = 1,\dots\}$.
Specifically, $\kappa_{n,m}$ is defined as the smallest positive $\kappa$ such that
\begin{equation}
\label{eqn:q1}
\begin{aligned}
\uni b^3  \left[
\frac{1}{n(m-1)} \sum_{l,h=1}^{\infty} \min\left\lbrace \kappa^2, \mu_l\mu_h\right\rbrace
\right]^{1/2} \leq  \kappa^2,\\
\quad 32\uni b \left[
\frac{1}{n(m-1)} \sum_{l,h=1}^{\infty} \min\left\lbrace \kappa^2/b^2, \mu_l\mu_h\right\rbrace
\right]^{1/2}  \leq    \kappa^2,
\end{aligned}
\end{equation}
where $c$ is  a universal constant,
and $\eta_{n,m}$ is defined as the smallest positive $\eta$ such that
\begin{align}
\label{eqn:q3}
\left(\frac{\vc}{nm} \sum_{l,h=1}^\infty \min\{\eta^2,\mu_l\mu_h\} + \frac{\eta^2}{n} \right)  ^{1/2}\leq \eta^2,
\end{align}
where $\vc$ is a constant depending on $\kc, b_X, b_\epsilon$.
The  existences  of $\kappa_{n,m}$ and $\eta_{n,m}$ are shown in the proof of Theorem \ref{thm_n,m_norm}.

\section*{Supplementary Material}

In the supplementary material related to this paper, we provide
proofs of our theoretical findings and additional simulation results.

\section*{Acknowledgement}
The research of Raymond K.~W.~Wong is partially supported by the U.S.~National Science Foundation under grants DMS-1806063, DMS-1711952 (subcontract) and CCF-1934904.
The research of Xiaoke Zhang is partially supported by the U.S.~National Science Foundation under grant DMS-1832046.
Portions of this research were conducted with high performance research computing resources provided by Texas A\&M University (\url{https://hprc.tamu.edu}).

\bibliographystyle{chicago}
\bibliography{plain_format.bbl}

\end{document}


\maketitle

\section{Proofs} \label{proof}

\subsection{Proof of Theorem 1}
\label{proof_re}
For any $\C\in \Gscr$ , we can decompose it into two orthogonal parts $\C_1$ and $\C_2$ such that $\C_1\in\Gscr(\mathcal{L}_{n,m})$ and $\C_2 \in \left( \Gscr(\mathcal{L}_{n,m})\right)^\bot $.  Since the loss function $\ell(\C)$ only depends on data, it suffices to show that $\Psi_0(\C_\squareun) \ge \Psi_0(\C_{1,\squareun})$ and $\Psi_k(\C_{(k)}) \ge \Psi_k(\C_{1,(k)})$ for $k=1,\ldots,p$. Below we follow two steps to prove this.

\medskip \noindent Step 1. Take $\Hscr(\mathcal{L}_{n,m}):= \bigotimes_{k=1}^p \mathcal{K}_k$.  Since we require $\C \in \mathcal{M}^+$, we first show that $\C_{1,\squareun}= \C_{1,\squareun}^\tp$ and $\langle \C_{1,\squareun} f, f\rangle_{\Hscr} \ge 0$ for any $f\in \Hscr $. Note that $\C_\squareun= \C_\squareun^\tp $, so $\C_\squareun =
(\C_{1,\squareun} + \C_{2,\squareun})/2 +
(\C_{1,\squareun}^\tp + \C_{2,\squareun}^\tp)/2$. As $\C_{1,\squareun}^\tp \in \Hscr(\mathcal{L}_{n,m}) \otimes \Hscr(\mathcal{L}_{n,m})$ and $\C_{2,\squareun}^\tp \in (\Hscr(\mathcal{L}_{n,m}) \otimes \Hscr(\mathcal{L}_{n,m}))^\bot$, we have $\C_1 =
(\C_{1,\squareun} + \C_{1,\squareun}^\tp)/2 $ and $\C_2=
(\C_{2,\squareun} + \C_{2,\squareun}^\tp)/2 $.  Thus $\C_{1,\squareun} = \C_{1,\squareun}^\tp$ and $\C_{2,\squareun}=\C_{2,\squareun}^\tp$.

By the definition of $\C_2$, $\langle \C_{2,\squareun} g,g\rangle _{\Hscr}=0$ for any $g \in \Hscr(\mathcal{L}_{n,m})$, so we have
$$0\leq \langle \C_\squareun g,g\rangle_{\Hscr} = \langle  \C_{1,\squareun} g,g\rangle_{\Hscr} + \langle  \C_{2,\squareun} g,g\rangle_{\Hscr} = \langle \C_{1,\squareun}g,g\rangle_{\Hscr}.$$
Moreover, the definition of $\C_1$ leads to $\langle  \C_{1,\squareun} g,g\rangle_{\Hscr} = 0$ for any $g\in (\Hscr(\mathcal{L}_{n,m}))^\bot$. Hence
$\langle \C_{1,\squareun} f, f\rangle_{\Hscr} \ge 0$ for any $f\in \Hscr $.

\medskip \noindent Step 2.     Next we show that for all $k$, $\lambda_k(\C_\squareun) \ge \lambda_k(\C_{1,\squareun})$ and $\lambda_k(\C_{(j)}) \ge \lambda_k(\C_{1,(j)}) $ with $j=1,\ldots,p$. Let $P_{\Hscr(\mathcal{L}_{n,m})}$ be the projection operator to space $\Hscr(\mathcal{L}_{n,m})$ and denote the adjoint operator of $A$ by $A^*$.  Then we have
\begin{align*}
\lambda_k(\C_{1,\squareun}) &=  \lambda_k\left( P_{\Hscr(\mathcal{L}_{n,m})}\C_\squareun P_{\Hscr(\mathcal{L}_{n,m})}\right) \\
& \leq  \lambda_k\left( \C_\squareun P_{\Hscr(\mathcal{L}_{n,m})}\right)  = \lambda_k\left( P_{\Hscr(\mathcal{L}_{n,m})}\C_\squareun^*\right)
\leq  \lambda_k\left( \C_\squareun^*\right) = \lambda_k(\C_\squareun).
\end{align*}
Let $P_{\mathcal{K}_j}$ denote the projection operator to space $\mathcal{K}_j$ and $P_{\mathcal{K}_{-j}}$ as the projection operator to space $\bigotimes_{k=1,k\neq j}^{2p} \mathcal{K}_k$ where $\mathcal{K}_{p+k} = \mathcal{K}_k$, $j=1,\ldots,p$. Then
\begin{multline*}
\lambda_k(\C_{1,{(j)}})  =  \lambda_k\left( P_{\mathcal{K}_j} \C_{(j)} P_{\mathcal{K}_{-j}}\right)  \leq  \lambda_k\left(\C_{(j)} P_{\mathcal{K}_{-j}} \right)
= \lambda_k\left( P_{\mathcal{K}_{-j}} \C_{(j)} ^*\right)  \leq  \lambda_k\left( \C_{(j)}^*\right) = \lambda_k\left( \C_{(j)}\right) .
\end{multline*}
Therefore,
$\Psi_0(\C_\squareun) \ge \Psi_0(\C_{1,\squareun})$ and $\Psi_k(\C_{(k)}) \ge \Psi_k(\C_{1,(k)})$ for $k=1,\ldots,p$.

\subsection{Proofs of Theorems 2,  3 and Corollary 1}
For notational simplicity,
we do not adopt different notations for the fully folded and squarely unfolded
versions of operators (functions) in this section.

Write $\Delta = \hat \C - \C_0$
and $\error (\bm T_{ij}, \bm T_{ij'}) = ( X_i(\bm T_{ij}) + \epsilon_{ij})( X_i(\bm T_{ij'}) +
\epsilon_{ij'}) - \C_0(\bm T_{ij},\bm T_{ij'}) $.
From (7),
we obtain the following basic inequality:
\begin{align}
\label{basic_inequality_convex}
\|\Delta\|_{n,m}^2 + \lambda I(\hat{\C}) &\leq 2 \langle  \error, \Delta \rangle _{n,m}+  \lambda I(\C_0).
\end{align}
The term $\langle \error, \Delta \rangle_{n,m}$ involved in \eqref{basic_inequality_convex}
plays a crucial role in the subsequent asymptotic analysis, so we will focus on this term first.

Consider $\Gscr_s = \{\left( \C - \C_0\right) /{\left\lbrace  I(\C)+I(\C_0)\right\rbrace }: \C \in \Gscr \}$ . To bound $\langle \error, \Delta \rangle_{n,m}$,   we start with controlling  $\sup_{g\in\Gscr_s}\langle \error , g \rangle_{n,m}$.
For any $g\in \Gscr_s$,  there exists a $\C\in\mathcal{G}$ such that $g= ( \C - \C_0) /\left\lbrace  I(\C)+I(\C_0)\right\rbrace  $. When $\C = \C_0$, $\|g\|_\Gscr=0$.
Otherwise,
$$
\|g\|_{\Gscr}= \left\|\frac{\C-\C_0}{I(\C)+I(\C_0) }\right\|_{\Gscr} \leq \frac{\|\C-\C_0\|_{\Gscr}}{I(\C - \C_0)} \leq \frac{\|\C-\C_0\|_{\Gscr}}{\|\C-\C_0\|_{\Gscr}}= 1,
$$
where the second inequality is due to that $I(\C) \ge \|\C\|_{\Gscr}$ for any
$\C\in \Gscr$, and $\| \cdot \|_\Gscr$ is Hilbert–Schmidt norm of RKHS $\Gscr$.
Take $\Gscr' =  \{g\in \Gscr: \|g\|_{\Gscr}\leq 1\}$.  From the above,
one can easily see that $\Gscr_s \subseteq \Gscr'$, and hence $\sup_{g\in\Gscr_s}\langle \error , g \rangle_{n,m} \leq \sup_{g\in\Gscr'}\langle \error , g \rangle_{n,m}$ for any $\error$.  In the later part of our analysis, we will bound $\sup_{g\in\Gscr'}\langle \error , g \rangle_{n,m}$ to control $\sup_{g\in\Gscr_s}\langle \error , g \rangle_{n,m} $.

First, we note that the functions residing in $\mathcal{G}'$ are bounded: For any $g \in \Gscr'$,  by the property of reproducing kernel,
\begin{align*}
\sup_{g\in \Gscr'} |g|_{\infty} \leq
\sup_{(x_1,...,x_{2p})\in [0,1]^{2p}} K ((x_1,...,x_{2p}),(x_1,...,x_{2p})) \leq \kc.
\end{align*}
Next we recall the definition of the sub-exponential norm of a random variable.
\begin{definition}
	For a random variable X, its sub-exponential norm is defined as
	\begin{align*}
	\|X\|_{\psi_1}=\inf\{\lambda  >0 : \E (\exp(|X|/\lambda))\leq 2\}.
	\end{align*}
	If $\|X\|_{\psi_1}< \infty$, then we call $X$ a sub-exponential random variable.
\end{definition}

Recall that  $\mathcal{L}_{n,m} = \lbrace T_{ijk}: i=1,...,n; j=1,...,m; k=1,...,p \rbrace $. We write  $\error_{ijj'} = \error (\bm T_{ij}, \bm T_{ij'})$.  For random variables $A$ and $B$, we denote by $\|A\mid B\|_{\psi_1}$ the sub-exponential norm of the random variable $A$ conditional on $B$.
The notation naturally extends to the case when $B$ is a random vector or a set of random variables.
By Lemma 3 in \citet{Wong-Zhang19}, we can see that conditioned on $\mathcal{L}_{n,m}$,
$\error_{ijj'}$ are  sub-exponential random variables.
Moreover, there exists a constant
$\sn$, depending on $b_X$ and $b_\epsilon$,
such that  $\|\error_{ijj'} \mid \mathcal{L}_{n,m}\|_{\psi_1} \leq \sn^2$.

Next we introduce the following random variables:
\begin{align*}
&\hat Z_{n,m}(\error, t; \Gscr'):=
\sup_{\{g\in\Gscr': \|g\|_{n,m}\leq t\}}
\left|\frac{1}{nm(m-1)}\sum_{i=1}^n \sum_{j\neq j'}^{m} \error_{ijj'}g (\bm T_{ij}, \bm T_{ij'})\right|,\\
&\tilde Z_{n,m}(\error, t; \Gscr'):=
\sup_{\{g\in\mathcal{G}': \|g\|_{2} \leq t\}}
\left|\frac{1}{nm(m-1)}\sum_{i=1}^n \sum_{j\neq j'}^{m} \error_{ijj'}g (\bm T_{ij}, \bm T_{ij'})\right|.
\end{align*}
Our immediate goal is to bound $\hat Z_{n,m}(\error, t; \Gscr')$, which will be achieved by bounding $\tilde Z_{n,m}(\error, t; \Gscr')$. We start with its expectation.
Without loss of generality, we use $\uni$ to denote all the universal constants.
\begin{lemma}
	\label{variance_control}
	There exists a constant $\vc>0$, depending on $\sn$ and $\Kf$, such that
	\begin{align}
	\E\left[\left\lbrace \tilde Z_{n,m}(\error, t; \Gscr')\right\rbrace^2\right] \leq \vc \left( \frac{1}{nm}\sum_{l,h=1}^\infty \min\{t^2,\mu_l\mu_h\} + \frac{t^2}{n} \right).
	\end{align}

\end{lemma}

\begin{proof}
	A majority of the proof resembles that of Lemma 42 in  \citet{mendelson2002geometric}, with additional arguments
	developed to control an important expectation term.
	Since the sample field of $X$ resides in $\Hscr$, we can decompose $X(\bm t) = \sum_{h=1}^{\infty} \zeta_h \phi_h(\bm t)$ where $\E  (\zeta_h\zeta_{h'}) = \E  \left\lbrace  \C_0(\bm T_{ij},\bm T_{ij'}) \phi_h(\bm T_{ij})\phi_h'(\bm T_{ij'})\right\rbrace $.
	For every $ \bm s,\bm t\in [0,1]^{p}$, write $\Phi(\bm s,\bm t) = \left(
	\sqrt{\mu_l \mu_h} \phi_l(\bm s) \phi_h(\bm t)\right)_{l,h=1}^{\infty} $. For
	two squarely summable sequences $a = \left\lbrace a_{lh}\right\rbrace
	_{l,h=1}^\infty$ and $b = \left\lbrace b_{lh}\right\rbrace _{l,h=1}^\infty$,
	define their inner product and the 2-norm in the following:
	$\langle a, b \rangle =\sum_{l,h=1}^{\infty} a_{lh} b_{lh} $ and $\|a\|_2 = (  \sum_{l,h=1}^{\infty}a^2_{lh}) ^{1/2} $.
	One can show that
	$$
	\Gscr' = \left\lbrace g(\cdot,\star) = \left\langle \beta, \, \Phi(\cdot,\star)\right\rangle :\|\beta\|_2\leq 1\right\rbrace.
	$$
	Let $\mathcal{B}(t) = \left\lbrace \beta : \|g\|_2\leq t\right\rbrace $. It follows that $g\in \Gscr'\cap \mathcal{B}(t)$  if and only if $\beta$ belongs to set  $\Omega = \{\beta : \sum_{l,h=1}^{\infty} \beta_{lh}^2 (\mu_l\mu_h) \leq t^2, \sum_{l,h=1}^{\infty} \beta_{lh}^2  \leq 1\}$.
	Let $\Xi = \lbrace \beta : \sum_{l,h=1}^{\infty} \beta_{lh}^2 \nu_{lh}\leq 1
	\rbrace  $, where $\nu_{lh}=( \min \lbrace 1, {t^2}/{\mu_l\mu_h} \rbrace )
	^{-1}$. We can see that $\Xi \subset  \Omega \subset \sqrt{2}\Xi$,
	which implies
	$$\E \left( \tilde Z_{n,m}(\omega, t; \Gscr')\right) ^2 \asymp \frac{1}{n^2m^2(m-1)^2} \E \sup_{\beta \in \Xi} \langle \beta ,\sum_{i=1}^{n} \sum_{j\neq j'}^{m} \error_{ijj'}  \Phi( \bm T_{ij}, \bm T_{ij'})\rangle^2.$$
	Next,
	\begin{align*}
	&\E \sup_{\beta \in \Xi} \left\langle \beta , \,\sum_{i=1}^{n} \sum_{j\neq j'}^{m} \error_{ijj'}  \Phi(\bm T_{ij},\bm T_{ij'})\right\rangle^2 \\
	&= \E \sup_{\beta \in \Xi} \left\langle \sum_{l,h=1}^{\infty}\sqrt{\nu_{lh}}\beta_{lh} , \,\sum_{l,h=1}^{\infty} \frac{\sqrt{\mu_l\mu_h}}{\sqrt{\nu_{lh}}}\sum_{i=1}^{n} \sum_{j\neq j'}^{m} \error_{ijj'}  \phi_l(\bm T_{ij})\phi_h(\bm T_{ij'}) \right\rangle^2\\
	&\leq \E\sum_{l,h=1}^{\infty} \frac{\mu_l \mu_h}{\nu_{lh}} \left\lbrace \sum_{i=1}^{n} \sum_{j\neq j'}^{m} \error_{ijj'} \phi_l(\bm T_{ij}) \phi_h(\bm T_{ij'})\right\rbrace ^2\\
	& = n \sum_{l,h=1}^{\infty} \frac{\mu_l \mu_h}{\nu_{lh}} \E  \left\lbrace\sum_{j\neq j'}^{m} \error_{1jj'} \phi_l(\bm T_{1j}) \phi_h(\bm T_{1j'}) \right\rbrace ^2.
	\end{align*}
	The last equality follows from the independence between different sample fields and observed locations,  combined with the fact that $\E(\error_{ijj'}\mid\mathcal{L}_{n,m}) = 0$.

	It remains to bound $\E \left\lbrace \sum_{j\neq j'}^{m} \error_{1jj'} \phi_l(\bm T_{1j}) \phi_h(\bm T_{1j'})\right\rbrace ^2 $.
	Write
	$$
	U_{jj'kk'} = \error_{1jj'} \error_{1kk'}\phi_l(\bm T_{1j}) \phi_h(\bm T_{1j'}) \phi_l(\bm T_{1k}) \phi_h(\bm T_{1k'}).
	$$
	When $j=k$ and $j'=k'$,
	\begin{align*}
	U_{jj'jj'}& = \E \error_{1jj'}^2 \phi^2_l(\bm T_{1j}) \phi^2_h(\bm T_{1j'}) = \E \left[ \left\lbrace \E(\error_{1jj'}^2\mid \mathcal{L}_{n,m})\right\rbrace \phi^2_l(\bm T_{1j}) \phi^2_h(\bm T_{1j'})\right]  \\
	&\leq \uni \sn^2 \E \left\lbrace \phi^2_l(\bm T_{1j}) \phi^2_h(\bm T_{1j'}) \right\rbrace  =\uni  \sn ^2,
	\end{align*}
	where the inequality follows from  the property of sub-exponential random variables and $\uni $ is a universal constant. When $j=k$ and $j'\neq k'$,
	\begin{align*}
	U_{jj'jk'}&= \E  \left\lbrace\error_{1jj'} \error_{1jk'}\phi_l^2(\bm T_{1j}) \phi_h(\bm T_{1j'}) \phi_h(\bm T_{1k'})\right\rbrace \\
	&\leq \E \left[  \left\lbrace  \E(\error_{1jj'}\mid  \mathcal{L}_{n,m})^2 \E(\error_{1jk'}\mid  \mathcal{L}_{n,m})^2\right\rbrace ^{1/2} \phi_l^2(\bm T_{1j}) \phi_h(\bm T_{1j'}) \phi_h(\bm T_{1k'})\right]   \\
	&\leq \uni \sn ^2  \E \left\lbrace \phi_l^2(\bm T_{1j}) \phi_h(\bm T_{1j'}) \phi_h(\bm T_{1k'})\right\rbrace  \leq \uni \sn ^2 \left\lbrace \E\phi^2_h(\bm T_{1j'}) \E\phi^2_h(\bm T_{1k'})\right\rbrace ^{1/2}\leq \uni\sn ^2.
	\end{align*}
	Similarly for  $j\neq k$ and $j' = k'$, $U_{jj'kj'} \leq \uni\sn ^2$.
	When $j\neq k$ and $j'\neq k'$,
	\begin{align*}
	U_{jj'kk'} &= \E\left\lbrace  \E \error_{1jj'} \phi_l(\bm T_{1j}) \phi_h(\bm T_{1j'}) \mid  X \right\rbrace  ^2\\
	& = \E \left[  \E  \left\lbrace  (X(\bm T_{1j})+ \epsilon_{1j})  (X(\bm T_{1j'})+ \epsilon_{1j'})-\C_0(\bm T_{1j},\bm T_{1j'})\right\rbrace   \phi_l(\bm T_{1j}) \phi_h(\bm T_{1j'})\mid X\right]  ^2\\
	&=\E \left[ \E \left\lbrace  X(\bm T_{1j})  X(\bm T_{1j'})\phi_l(\bm T_{1j}) \phi_h(\bm T_{1j'}) \mid X\right\rbrace  -\E \C_0(\bm T_{1j},\bm T_{1j'}) \phi_l(\bm T_{1j}) \phi_h(\bm T_{1j'})\right]    ^2\\
	&=\E \Bigg[   \E \left\lbrace \sum_{g=1}^{\infty} \zeta_g \phi_g(\bm T_{1j})\phi_l(\bm T_{1j})  \mid \{\zeta_g:g\ge 1\}\right\rbrace\\
	&\quad\times\E \left\lbrace \sum_{g=1}^{\infty} \zeta_g \phi_g(\bm T_{1j'})\phi_l(\bm T_{1j'})  \mid \{\zeta_g:g\ge 1\}\right\rbrace - \E \zeta_l \zeta_h\Bigg]   ^2\\
	&= \E(\zeta_l\zeta_h - \E\zeta_l \zeta_h)^2 \leq \E  \left(  \zeta^2_l\zeta^2_h\right).
	\end{align*}
	Putting together all these cases leads to
	\begin{align*}
	& \sum_{l,h=1}^{\infty} \frac{\mu_l \mu_h}{\nu_{lh}} \E\left\lbrace \sum_{j\neq j'}^{m} \error_{1jj'} \phi_l(\bm T_{1j}) \phi_h(\bm T_{1j'})\right\rbrace^2 \\
	&\leq  \sum_{l,h=1}^{\infty}\frac{\mu_l \mu_h}{\nu_{lh}} \left\lbrace  m(m-1) \uni \sn ^2 + 3m(m-1)(m-2) \uni  \sn ^2 + m(m-1)(m-2)(m-3)\E  \left(  \zeta^2_l\zeta^2_h\right)\right\rbrace  \\
	&\leq \uni  \left\lbrace   m^3\uni \sn ^2  \sum_{l,h=1}^{\infty} \frac{\mu_l \mu_h}{\nu_{lh}}  + m^4  \sum_{l,h=1}^{\infty} \frac{\mu_l \mu_h}{\nu_{lh}} \E  \left(  \zeta^2_l\zeta^2_h\right)\right\rbrace \\
	&\leq \uni \left\lbrace  m^3 \uni \sn ^2  \sum_{l,h=1}^{\infty} \min\{t^2, \mu_l\mu_h\} + m^4 t^2  \sum_{l,h=1}^{\infty}\E  \left(  \zeta^2_l\zeta^2_h\right)\right\rbrace .
	\end{align*}
	Since $\sum_{l,h=1}^{\infty}\E  \left(  \zeta^2_l\zeta^2_h\right) = \E(X^4(\bm T)) = \Kf < \infty$,
	\begin{align*}
	\E \left\lbrace \tilde Z_{n,m}(\error, t; \Gscr')\right\rbrace  ^2 \leq \vc \left( \frac{1}{nm}\sum_{l,h=1}^\infty \min\{t^2,\mu_l\mu_h\} + \frac{t^2}{n} \right).
	\end{align*}

\end{proof}

Next we derive the following concentration inequality for $\tilde Z_{n,m}(\error, t; \Gscr')$.
\begin{lemma}
	\label{adamzack}
	There exists a universal constant $\uni>1$ and a constant $\ac>0$ depending on $\kc$ and $\sn$, such that with probability at least $1-\exp(-c{nt^2}/{\log n})$,  we have
	$$\tilde Z_{n,m}(\error, t; \Gscr') \leq \uni \left\lbrace \E \tilde Z_{n,m}(\error, t; \Gscr') + \ac t^2 \right\rbrace. $$
\end{lemma}

\begin{proof}
	Write $\bm \error_i = \left\lbrace \error_{ijj'}: j=1,...,m\right\rbrace $, $\bm T_i = \left\lbrace \bm T_{ij}: j=1,...,m\right\rbrace $ and
	$$
	f(\bm \error_i ,\bm T_i) = \frac{1}{m(m-1)}\sum_{j\neq j'}^{m} \error_{ijj'} g(\bm T_{ij}, \bm T_{ij'}).
	$$
	Note that $	\E (f(\bm \error_i, \bm T_i)) = 0$.
	We adopt the Adamczak bound \citep[Theorem 4 in][]{adamczak2008tail, koltchinskii2011oracle}  to establish a concentration inequality for the unbounded class $\Fscr=\left\lbrace f: g \in \Gscr',  \|g\|_2 \leq t\right\rbrace $.
	To this end,
	we need to bound a variance term
	$\sigma^2(\mathcal{F}):=
	\sup_{f\in\mathcal{F}}
	\E(f^2(\bm \error_1,\bm T_1))$ and
	the sub-exponential norm of the envelope function $F$ of the class $\Fscr$.
	For the variance term,
	\begin{align*}
	&\sigma^2(\mathcal{F}):= \sup_{\|g\|_{\Gscr}\leq 1, \|g\|_2 \leq t}  \E f^2(\bm \error_1,\bm T_1))\\
	&=\frac{1}{m^2(m-1)^2}\sup_{\|g\|_{\Gscr}\leq 1, \|g\|_2 \leq t} \E \left\lbrace \sum_{j\neq j'}^{m} \error_{1jj'} g(\bm T_{1j},\bm T_{1j'})\right\rbrace ^2\\
	& = \frac{1}{m^2(m-1)^2} \sum_{j\neq j'}^{m}  \sum_{k\neq k'}^{m}\sup_{\|g\|_{\Gscr'}\leq 1, \|g\|_2 \leq t} \E(\E \error_{1jj'}\error_{1kk'}\mid \bm T_1)g(\bm T_{1j},\bm T_{1j'}) g(\bm T_{ik},\bm T_{ik'})\\
	& \leq \frac{\uni \sn^2 }{m^2(m-1)^2}\sum_{j\neq j'}^{m}  \sum_{k\neq k'}^{m}\sup_{\|g\|_{\Gscr}\leq 1, \|g\|_2 \leq t} \E g(\bm T_{1j},\bm T_{1j'}) g(\bm T_{ik},\bm T_{ik'})\\
	&\leq \frac{\uni \sigma_{\psi_1} ^2}{m^2(m-1)^2}\sum_{j\neq j'}^{m}  \sum_{k\neq k'}^{m}\sup_{\|g\|_{\Gscr}\leq 1, \|g\|_2 \leq t} \left\lbrace  \E g^2(\bm T_{1j},\bm T_{1j'}) \E g^2(\bm T_{ik},\bm T_{ik'})\right\rbrace  ^{1/2}\\
	&\leq \uni \sn ^2t^2.
	\end{align*}
	As for the envelope,
	\begin{align*}
	&\|\max_{i=1,...,n} F(\bm \error_i,\bm T_i) \|_{\psi_1} \leq  \uni \max_{i=1,...,n}\|F(\bm \error_i,\bm T_i) \|_{\psi_1} (\log n) \\
	&\leq \frac{\uni  \kc }{m(m-1)} \|\sum_{j\neq j'}^{m} \error_{ijj'} \|_{\psi_1}(\log n)
	\leq \uni  \kc  \sn ^2(\log n),
	\end{align*}
	where the first inequality comes from Theorm 4 of \citet{pisier1983some}  and the second inequality results from  $g(\bm T_{ij},\bm T_{ij'})\leq b$.
	The desired result then follows from Adamzack bound.
\end{proof}

By Lemmas \ref{variance_control} and \ref{adamzack}, we are able to bound $\tilde Z_{n,m}(\error, t; \Gscr')$.  Then, we relate $\hat Z_{n,m}(\error, t; \Gscr')$ with $\tilde Z_{n,m}(\error, t; \Gscr')$ by Lemma \ref{norm_control} below.
Recall that $\kappa_{n,m}$ is
the smallest positive real number $\kappa$ that fulfills the following inequalities
\begin{align}
\label{eqn:Q1}
\uni b^3 Q(\kappa/b) \leq  \kappa^2,\\
\label{eqn:Q2}
32\uni b Q(\kappa) \leq    \kappa^2,
\end{align}
where $c$ is  an universal constant that we do not specify and
$$
Q(\kappa)= \left[
\frac{1}{n(m-1)} \sum_{l,h=1}^{\infty} \min\left\lbrace \kappa^2, \mu_l\mu_h\right\rbrace
\right]^{1/2}.
$$
Note that $Q(\kappa)/\kappa\rightarrow\infty$ as $\kappa\rightarrow 0$.
Also, $Q(\kappa)/\kappa$ is non-increasing in $\kappa$.
Dividing both sides in \eqref{eqn:Q1}  and \eqref{eqn:Q2} by $\kappa$,
the resulting right hand side is an identity function, which is continuous, strictly increasing and is zero when $\kappa=0$.
Therefore $\kappa_{n,m}$ exists.

\begin{lemma}
	\label{norm_control}
	We assume $t \ge \kappa_{n,m}$ for all the following cases.  For any $g \in \Gscr'$,
	there exist constants $\nc_1, \nc_2 > 2$,
	both depending on $\kc$, such that
	\begin{align*}
	\left\lbrace \|g\|^2_{n,m} \leq t^2\right\rbrace \subseteq\left\lbrace \|g\|^2_2 \leq \nc_1 t^2 \right\rbrace,
	\end{align*}
	with probability at least $1-\exp(-\uni n m\kappa_{n,m}^2 + \log m)$, and
	\begin{align*}
	\left\lbrace \|g\|^2_{2} \leq t^2\right\rbrace \subseteq \left\lbrace \|g\|^2_{n,m} \leq \nc_2 t^2 \right\rbrace,
	\end{align*}
	with probability at least $1-\exp(-\uni n m\kappa_{n,m}^2 + \log m)$.
	Additionally,
	we have
	\begin{align*}
	\|g\|_2^2 - \|g\|_{n,m} ^2\leq \frac{1}{2} \|g\|_2^2,
	\end{align*}
	holds for all $g\in\mathcal{G}'$ such that $\|g\|_2^2>t^2$,
	with probability at least 	$1-\exp(-\pc nm t^2+ {\log m})$ where $\pc$ is a constant depending on $b$.
\end{lemma}

\begin{proof}
	For $1\le j,j'\le m$, we call $(j,j')$ a pair formed by individuals $j$ and $j'$. When $m$ is even,  by Lemma \ref{grouping}, we are able to partition the collection $\mathcal{P}=\left\lbrace (j,j'):1\leq j< j' \leq m \right\rbrace $ into $(m-1)$ groups $G_1,...,G_{m-1}$, such that
	$G_k\cap G_{k'}=\emptyset$ for $k\neq k'$,
	$\mathcal{P}=\bigcup_{k=1}^{m-1} G_k$,
	$\mathrm{card}(G_k)=m/2$ for all $k$,
	and
	$\mathrm{card}(\{(j,j')\in G_k: \mbox{$j=\tilde{j}$ or $j'=\tilde{j}$}\})=1$ for all $\tilde{j}$ and $k$
	(i.e., no individual occurs more than one time within a group),
	where $\mathrm{card}(A)$ denotes the cardinality of a set $A$.
	Therefore it is easy to see that the location pairs in $\{\left(T_{ij} , T_{ij'} \right): (j,j')\in G_k\} $ are independent for any fixed $k$.
	As an illustration, suppose $m=4$. Following the construction rule in Lemma \ref{grouping}, we obtain three groups $G_1 = \left\lbrace (1,4), (2,3)\right\rbrace $, $G_2 = \left\lbrace (1,2), (3,4)\right\rbrace $ and $G_3 = \left\lbrace (1,3), (2,4)\right\rbrace $.

	Consider the case when $m$ is even. 	Take $f_{G_k}(\bm T) = \frac{2}{nm} \sum_{i=1}^n\sum_{(j,j')\in G_k} g^2(\bm T_{ij},\bm T_{ij'})$, $k=1,...,m-1$.
	Note that the $nm/2$ summands $g^2(\bm{T}_{ij}, \bm{T}_{ij'})$
	in $f_{G_k}(\bm{T})$
	all have expectation $\|g\|_2^2$, and are independent due to the above
	grouping property.
	To relate $\|g\|_2^2$ and $f_{G_k}(\bm{T})$, we can apply  Theorem 3.3  in \cite{bartlett2005local}.

	Take $R_{n,m} (t;G_k, \Gscr') = \frac{2}{nm}\sup_{\{g \in \Gscr': \|g\|_2\leq t\}}| \sum_{i=1}^n\sum_{(j,j')\in G_k} \sigma_{ijj'} g^2(\bm T_{ij},\bm T_{ij'})|$ to be the corresponding empirical local Rademacher complexity.  By  the well-known contraction inequality and Lemma 42 in \cite{mendelson2002geometric}, it is simple to show that with some universal constant $\uni$,
	\begin{align*}
	\E R_{n,m} (t;G_k, \Gscr') &\leq 2b\frac{2}{nm}\E\left\{\sup_{g \in \Gscr', \|g\|_2\leq t}\left| \sum_{(j,j')\in G_k} \sigma_{ijj'} g(\bm T_{ij},\bm T_{ij'})\right|\right\}\\
	&\leq \uni b \left( \frac{1}{nm}\sum_{l,h=1}^{\infty} \min \{ t^2, \mu_l\mu_h\}\right) ^{1/2} \le \uni b Q(t)
	\end{align*}
	Note that for $(j,j')\in \mathcal{G}_k$,
	\begin{align*}
	\text{Var}\{g^2(\bm T_{ij},\bm T_{ij'}) \} \leq  \E\{g^4(\bm T_{ij},\bm T_{ij'})\}\leq
	b^2\|g\|_2^2\le b^2t^2.
	\end{align*}
	In Theorem 3.3 in \cite{bartlett2005local},
	we can take $T(g) =b^2 \|g\|_2^2$, $B = b^2$ and
	$\psi(r)=\uni b^3 Q(r^{1/2}/b)$.
	We then verify a condition in Theorem 3.3 in \cite{bartlett2005local}.
	For any $t>0$,
	\begin{align*}
	b^2 \E R_{n,m} (t;G_k, \Gscr') &= \frac{2b^2}{nm}\E\left\{\sup_{g \in
		\Gscr', T(g)\leq b^2t^2}
	\left| \sum_{(j,j')\in G_k} \sigma_{ijj'} g^2(\bm T_{ij},\bm T_{ij'})\right|\right\}
	\leq cb^3 Q(t),
	\end{align*}
	where the desired condition  follows from
	taking $r = b^2t^2$.
	From the definition \eqref{eqn:Q1} of $\kappa_{n,m}$,   we can see that $\kappa^2_{n,m} $ is larger than the fixed point of $\psi$ (i.e., the solution of $\psi(r) = r$).
	Theorem 3.3 in \cite{bartlett2005local} shows that
	\begin{align*}
	\|g\|^2_2 \leq 2f_{G_k}(\bm T)  + \frac{1408}{b^2} \kappa^2_{n,m} + 2(11b^2 + 52b^2)\kappa^2_{n,m} = 2 f_{G_k}(\bm T) +
	\left(\frac{1408}{b^2}+ 126b^2\right)\kappa_{n,m}^2,
	\end{align*}
	holds for all $g\in\mathcal{G}'$,
	with probability at least $1-\exp(-nm\kappa^2_{n,m})$.
	Also,
	\begin{align*}
	f_{G_k}(\bm T) \leq 2\|g\|^2_{2} + \frac{704}{b^2} \kappa^2_{n,m} +
	2(11b^2 + 26b^2) \kappa_{n,m}^2 = 2\|g\|_2^2
	+ \left( \frac{704}{b^2} + 74b^2 \right)\kappa_{n,m}^2,
	\end{align*}
	holds for all $g\in\mathcal{G}'$,
	with probability at least $1-\exp(-nm\kappa_{n,m}^2)$.

	Recall that $\|g\|_{n,m}^2 =\frac{1}{m-1} \sum_{i=1}^{m-1} f_{G_k}(\bm T)$.
	We proceed by taking union bounds of the probability statements derived above, over $f_{G_1},...,f_{G_{m-1}}$.
	If $t\ge \kappa_{n,m}$,
	\begin{align*}
	\|g\|^2_2 \leq 2\|g\|_{n,m}^2
	+\left(\frac{1408}{b^2}+ 126b^2\right)\kappa_{n,m}^2
	\leq \nc_1 t^2,
	\end{align*}
	holds for all $g\in\mathcal{G}'$ such that $\|g\|_{n,m}^2\le t^2$,
	with probability at least $1-(m-1)\exp(-nm\kappa^2_{n,m})$.
	Also, if $t\ge \kappa_{n,m}$,
	\begin{align*}
	\|g\|_{n,m}^2 \leq 2\|g\|^2_{2}
	+ \left( \frac{704}{b^2} + 74b^2 \right)\kappa_{n,m}^2
	\leq \nc_2 t^2,
	\end{align*}
	holds for all $g\in\mathcal{G}'$ such that $\|g\|_2^2\leq t^2$,
	with probability at least $1-(m-1)\exp(-nm\kappa_{n,m}^2)$.
	Here $\nc_1,\nc_2>2$ are constants that depend on $b$.

	Now, we focus on $\|g\|_2^2 > t^2$.
	By applying Theorem 2.1 in \cite{bartlett2005local},
	we obtain the following inequality
	\begin{align*}
	\|g\|_2^2 - f_{G_k}(\bm T)\leq 0.5 \|g\|_2^2,
	\end{align*}
	holds for all $g\in\mathcal{G}'$ such that $\|g\|_2^2>t^2$,
	with probability at least $1-\exp(-(mn/64b^2) t^2$.
	Take a union bound over $(m-1)$ groups, we will have
	\begin{align*}
	\|g\|_2^2 - \|g\|_{n,m}^2\leq 0.5 \|g\|_2^2,
	\end{align*}
	holds for all $g\in\mathcal{G}'$ such that $\|g\|_2^2>t^2$,
	with probability at least $1-(m-1)\exp(-(mn/64b^2) t ^2)$.

	When $m$ is odd, $\left\lbrace(j,j') :1\leq j< j'\leq m-1 \right\rbrace $ can be decomposed into $(m-2)$ groups ($G_1,\dots,G_{m-2}$) as described before,
	since $m-1$ is even.
	The remaining pairs are $\left\lbrace (j,m): j=1,2,...,m-1\right\rbrace $
	which are not independent.
	\begin{align}
	\|g\|_{n,m}^2 &= \frac{m-2}{m}\frac{1}{(m-2)}\left\{\sum_{k=1}^{m-2} \sum_{i=1}^n\frac{2}{n(m-1)} \sum_{(j,j')\in G_k} g^2(\bm T_{ij}, \bm T_{ij'}) \right\}+ \frac{2}{m(m-1)}\sum_{j=1}^{m-1} \frac{1}{n} \sum_{i=1}^n g^2(\bm T_{ij},\bm T_{im})
	\label{eqn:lem4odd}
	\end{align}
	In the odd-$m$ setting, we define $f_{G_k}(\bm T)=\frac{2}{n(m-1)}\sum_{i=1}^n \sum_{(j,j')\in G_k} g^2(\bm T_{ij}, \bm T_{ij'})$. We can apply the similar arguments derived for the even case (with $m$ replaced by $m-1$).
	Therefore, we focus on the new term, which is the second term in \eqref{eqn:lem4odd}.
	First, we study $V_j(\bm T) = \frac{1}{n}\sum_{i=1}^n g^2(\bm T_{ij},\bm T_{im})$
	for a fixed $1\le j \le m-1$.
	Note that $\E\{g^2(\bm{T}_{ij}, \bm{T}_{im})\} =\|g\|_2^2$
	and the summands in $V_j(\bm{T})$ are independent.

	We still apply Theorem 3.3 in \citet{bartlett2005local}.  The local Rademacher complexity becomes
	$$R(t;\Gscr') = \E\left\{\frac{1}{n} \sup_{g\in \Gscr', \|g\|_2\leq t} \sum_{i=1}^n g^2(\bm T_{ij},\bm T_{im})\right\}  \leq \uni b \left( \frac{1}{n}\sum_{l,h=1}^{\infty} \min \{ t^2, \mu_l\mu_h\}\right) ^{1/2} .$$
	Take $\kappa'_{n} $ to be the smallest positive real number $\kappa$ that satisfies
	\begin{align*}
	\uni b^3 \left( \frac{1}{n}\sum_{l,h=1}^{\infty} \min \{ (\kappa/b)^2, \mu_l\mu_h\}\right) ^{1/2}\leq \kappa^2.
	\end{align*}
	By Theorem 3.3 in \citet{bartlett2005local}, it can be shown that
	\begin{align*}
	\|g\|^2_2 &\leq 2V_j(\bm T)  + \frac{1408}{b^2} \kappa'^{2}_{n}+ 2(11b^2 + 52b^2)m\kappa_{n,m}^2\\
	\|g\|^2_2 /m & \leq 2V_j(\bm T)/m   + \frac{1408}{b^2} \kappa'^{2}_{n}/m + 2(11b^2 + 52b^2)\kappa_{n,m}^2
	\end{align*}
	holds for all $g\in\mathcal{G}'$,
	with probability at least $1-\exp(-nm\kappa_{n,m}^2)$.
	Also,
	\begin{align*}
	V_j(\bm T)/m \leq 2\|g\|^2_{2}/m + \frac{704}{b^2} \kappa'^{2}_{n} /m + 2(11(b^2) + 26b^2) \kappa_{n,m}^2,
	\end{align*}
	holds for all $g\in\mathcal{G}'$, with probability at least $1-\exp(-nm\kappa_{n,m}^2)$.

	Now, we take a union bound,
	and then combine it with the result for the first term in \eqref{eqn:lem4odd}. Since $ \kappa'^{2}_{n}/m \leq \kappa^2_{n,m}$, we derive the following with assumption $t \ge \kappa_{n,m}^2$:

	\begin{align*}
	\|g\|^2_2 \leq 2\|g\|_{n,m}^2  + \frac{1408}{b^2} \kappa^2_{n,m}\left(\frac{m-2}{m} + 2\right) + 2(11b^2 + 52b^2)\kappa^2_{n,m}\left(\frac{m-2}{m} + 2\right)  \leq \nc_1 t^2
	\end{align*}
	holds for all $g\in\mathcal{G}'$ such that $\|g\|_2^2\leq t^2$, with probability at least $1-(m-2 + 2(m-1))\exp(-nm\kappa_{n,m}^2)$.

	\begin{align*}
	\|g\|_{n,m}^2 \leq 2\|g\|^2_{2} + \frac{704}{b^2} \kappa^2_{n,m}\left(\frac{m-2}{m} + 2\right)  + 2(11(b^2) + 26b^2) \left(\frac{m-2}{m} + 2\right) \kappa_{n,m}^2 \leq \nc_2 t^2
	\end{align*}
	holds for all $g\in\mathcal{G}'$ such that $\|g\|_2^2<t^2$, with probability at least $1-(m-2 + 2(m-1))\exp(-nm\kappa_{n,m}^2)$. Here $\nc_1$ and $\nc_2$ are some constants that depend on $b$.

	With similar argument, we will be able to derive for the odd case,
	\begin{align*}
	\|g\|_2^2 - \|g\|_{n,m}^2\leq 0.5 \|g\|_2^2,
	\end{align*}
	holds for all $g\in\mathcal{G}'$ such that $\|g\|_2^2>t^2$,
	with probability at least $1-(m-2 + 2(m-1))\exp(-\pc nm t^2)$ for some constant $\pc = \pc(1/b)$.

\end{proof}

With Lemmas \ref{variance_control}, \ref{adamzack} and \ref{norm_control},
we are now ready to prove Theorem 2. Recall the definition of $\eta_{n,m}$ and $\xi_{n,m}$.
The term $\eta_{n,m}$ is defined as the smallest positive value $\eta$ such that
\begin{align*}
\left(\frac{\vc}{nm} \sum_{l,h=1}^\infty \min\{\eta^2,\mu_l\mu_h\} + \frac{\eta^2}{n} \right)  ^{1/2}\leq \eta^2,
\end{align*}
where $\vc>0$ is a constant defined in Lemma \ref{variance_control}.
By similar arguments for the existence of $\kappa_{n,m}$,
we can show that $\eta_{n,m}$ exists.
By Lemma \ref{variance_control}, we can show that
$\E \tilde{Z}_{n,m}(\error,t;\Gscr')\le
\sqrt{\E [\{\tilde{Z}_{n,m}(\error,t;\Gscr')\}^2]}
\leq t^2$ for $t\ge \eta_{n,m}$.

Take $\xi_{n,m} = \min\left\lbrace  \max\left\lbrace \eta_{n,m}, \kappa_{n,m}\right\rbrace , \left( \frac{\log n}{n} \right)^{1/2} \right\rbrace $.  We include
the term $({\log n}/{n} ) ^{1/2}$ mainly
due to the unboundedness of $\{\error_{ijj'}\}$,
which leads to the application of Adamzack bound (Lemma \ref{adamzack})
instead of simpler forms of Talagrand's concentration inequality.

\begin{proof} [Proof for Theorem 2]
	First, we study the crucial term
	\[
	\hat Z_{n,m}(\error,\kc;\Gscr')=
	\sup_{\{g\in\mathcal{G}':\|g\|_{n,m}\leq b\}} \left|\frac{1}{nm(m-1)}\sum_{i=1}^n \sum_{j\neq j'} \error_{ijj'}g (\bm T_{ij}, \bm T_{ij'})\right|,
	\]
	which is bounded by
	the maximum of $\hat Z_{n,m}\left( \error, \xi_{n,m}; \Gscr' \right)$ and
	\begin{equation}
	\sup_{\{g\in\Gscr': \|g\|_{n,m}> \xi_{n,m}\}}\left|\frac{1}{nm(m-1)}\sum_{i=1}^n \sum_{j\neq j'} \error_{ijj'}g (\bm T_{ij}, \bm T_{ij'})\right|,
	\label{eqn:thm2second}
	\end{equation}
	so it suffices to study the rates of convergence of these two terms.

	For the rate of the maximum of $\hat Z_{n,m}\left( \error, \xi_{n,m}; \Gscr' \right)$,
	by Lemmas \ref{variance_control}, \ref{adamzack} and \ref{norm_control}, we can show that with probability at least $1-\exp(-\uni n\xi_{n,m}^2/\log n )$ for some universal constant $\uni$:
	\begin{align*}
	\hat Z_{n,m}(\error,\xi_{n,m};\Gscr') &\leq \tilde Z_{n,m} (\error, \sqrt{\nc_1} \xi_{n,m} ;\Gscr') \\
	&\leq \uni \left\lbrace \E \tilde Z_{n,m}(\error, \sqrt{\nc_1} \xi_{n,m} ;\Gscr') +  \ac \nc_1  \xi_{n,m}^2\right\rbrace  \\
	& \leq \uni\left\lbrace \nc_1\xi_{n,m}^2 + \ac\nc_1 \xi_{n,m}^2\right\rbrace \leq \czhat\xi_{n,m}^2,
	\end{align*}
	where $\czhat = cM_1(1+c_1)$ and,
	the first, second and last inequalities are due to Lemmas \ref{norm_control}, \ref{adamzack} and \ref{variance_control} respectively.

	For the rate of the second term in \eqref{eqn:thm2second},
	we first prove the following result.
	For any $r > \xi_{n,m}$, with probability at least $1-\exp(-\uni {n}\xi_{n,m}^2/\log n)$,
	we have
	\begin{align}
	\hat Z_{n,m}(\error,r ;\Gscr') &= \frac{r}{\xi_{n,m}} \sup_{\{g\in\mathcal{G}':\|g\|_{\Gscr}\leq \frac{\xi_{n,m}}{r}, \|g\|_{n,m} \leq \xi_{n,m}  \}}\left|\frac{1}{nm(m-1)}\sum_{i=1}^n \sum_{j\neq j'} \error_{ijj'}g (\bm T_{ij}, \bm T_{ij'})\right| \nonumber\\
	& \leq \frac{r}{\xi_{n,m}}\hat {Z}_{n,m}(\error,\xi_{n,m};\Gscr')
	\leq \frac{r}{\xi_{n,m}} \czhat\xi_{n,m}^2 = \czhat r \xi_{n,m}. \label{eqn:thm2Zr}
	\end{align}

	For $b>\xi_{n,m}$, a direct application of the above result with $r=b$ does not provide the increment with respect to the empirical norm, and so we apply a peeling argument.

	Set $S_l := \{g\in\mathcal{G}':2^{l-1}\xi_{n,m}\leq \|g\|_{n,m} \leq 2^{l}\xi_{n,m}\}$, $l=1,\ldots,L$,  where $L= \log_2(b/\xi_{n,m})$.
	\begin{align*}
	&\pr \left( \sup_{\{g\in\mathcal{G}': \|g\|_{n,m} > \xi_{n,m}\}} \frac{\left|\frac{1}{nm(m-1)}\sum_{i=1}^n \sum_{j\neq j'} \error_{ijj'}g (\bm T_{ij}, \bm  T_{ij'})\right|}{\|g\|_{n,m}}> 2\czhat \xi_{n,m}\right) \\
	&\leq\sum_{l=1}^L \pr  \left( \sup_{g\in S_l}  \frac{\left|\frac{1}{nm(m-1)}\sum_{i=1}^n \sum_{j\neq j'} \error_{ijj'}g (\bm T_{ij}, \bm T_{ij'})\right|}{\|g\|_{n,m}}> 2\czhat \xi_{n,m}\right) \\
	&\leq \sum_{l=1}^L  \pr\left( \hat{Z}_{n,m} (\error, 2^l\xi_{n,m};\Gscr')> 2\czhat  2^{l-1}\xi^2_{n,m}\right) \\
	&\leq L \exp\left(-c \frac{n}{\log n} \xi_{n,m}^2\right)\\
	&\leq \exp\left(-c \frac{n}{\log n} \xi_{n,m}^2\right),
	\end{align*}
	where the second last inequality results from \eqref{eqn:thm2Zr} by taking
	$r=2^l\xi_{n,m}$, and the universal constant $c$ in the last two lines could be different.
	For the last inequality, as long as $ 0\leq (\log(\log (1/\xi_{n,m}))) / \{\frac{n}{\log n}\xi_{n,m}^2 \}\leq 1$
	such a universal constant $c$ exists.

	Therefore, we have
	\begin{align*}
	\langle\error,g\rangle_{n,m} &\leq \czhat(\xi_{n,m}^2 + 2\|g\|_{n,m} \xi_{n,m}),
	\end{align*}
	for every $g \in \Gscr_s \subset \Gscr'$,
	with probability at least $1-\exp(-\uni n\xi_{n,m}^2/\log n) $.
	With the same probability, we have
	\begin{align}
	\label{inner1}
	\langle \error, \Delta\rangle _{n,m} \leq  \czhat \xi_{n,m}^2\left\lbrace  I(\hat \C ) +I( \C_0) \right\rbrace +  2\czhat \xi_{n,m}  \|\Delta\|_{n,m}.
	\end{align}
	In below, we condition on the event \eqref{inner1}.
	From the basic inequality \eqref{basic_inequality_convex}, with $\lambda = c_\lambda \xi_{n,m}^2 $ such that $c_\lambda > 2\czhat $,
	\begin{align*}
	\|\Delta\|_{n,m}^2 & \leq 2 \langle \error, \Delta \rangle_{n,m} +  \lambda (I(\C_0)- I(\hat \C)), \\
	\|\Delta\|_{n,m}^2 & \leq 2\lambda I(\C_0)+  4\czhat \xi_{n,m}\|\Delta\|_{n,m}.
	\end{align*}
	Then we have
	\begin{align*}
	\|\Delta\|_{n,m} \leq \left\lbrace 2c_\lambda I(\C_0)\right\rbrace ^{\frac{1}{2}}\xi_{n,m} +   4\czhat  \xi_{n,m}
	\end{align*}
	and the proof is complete by taking $L_1 =2 \czhat$.
\end{proof}

Next, we are ready to bound the $L^2$ norm $\|\Delta\|_{2}$ for $\Delta=\hat \C - \C_0$ obtained by (7).

\begin{proof}[Proof of Theorem 3]
	From Lemma \ref{norm_control}, we can see that $\|g\|_2^2 \leq 2\|g\|_{n,m}^2  + \xi^2_{n,m}$, for all $g \in \Gscr'$ with probability at least $1-\exp(-\pc \xi_{n,m}^2)$ for some universal constant $\pc=\pc(1/b)$. So with the same probability
	we have $\|\Delta\|_2 \leq 2^{1/2} \|\Delta\|_{n,m}+ \xi_{n,m} \left\lbrace  I(\hat \C) + I(\C_0)\right\rbrace$.  In terms of Lemma \ref{regularization bound}, we are able to  bound the regularization term $I(\hat \C)$ by a constant $L_2$, so finally we get
	\begin{align*}
	\|\Delta\|_2 &\leq 2^{\frac{1}{2}}\left[  \left\lbrace 2 c_\lambda I(\C_0)\right\rbrace ^{\frac{1}{2}} + 4\czhat \right]  \xi_{n,m} + \left\lbrace R_2 + I(\C_0)\right\rbrace  \xi_{n,m}\\
	&\leq \left[  2\left\lbrace c_\lambda I(\C_0)\right\rbrace ^{\frac{1}{2}}+ 4(2)^{\frac{1}{2}} \czhat  + R_2 + I(\C_0) \right]\xi_{n,m} .
	\end{align*}
	By taking $L_2=  4(2)^{1/2} \czhat  + R_2 + I(\C_0)$, the proof is complete.
\end{proof}

\begin{proof}[Proof of Corollary 1]
	By Lemma \ref{tensor_sequence}, the tensor product eigenvalue sequence has decay $\mu_l\asymp  (l^{-2\alpha} (\log l)^{2\alpha(2p-1)})$ as $l\rightarrow \infty$.

	By the definitions of $\kappa_{n,m}$ and $\eta_{n,m}$, when $m=\mathcal{O} \left( n^{1/(2\alpha)}(\log n)^{2p-2-1/(2\alpha)} \right)$, they are all of the same order, and so is $\xi_{n,m}$. By Lemma \ref{order}, we can see that  \\$\xi_{n,m}\asymp (nm)^{{2\alpha }/{(1+2\alpha )}} (\log nm)^{{2\alpha(2p-1)}/{(2\alpha +1) }}$. When $n^{1/(2\alpha)}(\log n)^{2p-2-\frac{1}{2\alpha}} = \mathcal{O}(m)$, $\log n/n$ will be the dominant term. From Theorems 2 and 3, we can see that $\|\hat{\C}-\C_0\|_{n,m}^2$ and $\|\hat{\C}-\C_0\|_{2}^2$ are both of the same order.  Overall, we have

	$$\|\hat \C - \C_0\|^2_{n,m}, \|\hat \C - \C_0\|^2_{2}= \bigOp \left( (nm)^{-\frac{2\alpha }{1+2\alpha }}(\log nm)^{\frac{2\alpha(2p-1)}{2\alpha +1 }} + \frac{\log n}{n}\right). $$
\end{proof}

\subsection{Auxiliary Lemmas}

\begin{lemma}
	\label{grouping}
	When $m$ is even, we can decompose any collection of individual index pairs $\left\lbrace (j,j') : 1\leq j<j'\leq m\right\rbrace $ into $(m-1)$ groups such that within each group, there are $m/2$ pairs and no repeated individuals.
\end{lemma}

\begin{proof}
	First, we consider to construct a matrix $\bm G\in \mathbb{R}^{m\times m}$ that satisfies following conditions:
	1. All the diagonal entries are zero;\\
	2. Every row and every column is a permutation of sequence $\left\lbrace 0,1,2,...,(m-1)\right\rbrace $;\\
	3. It is symmetric.

	To begin with,  we  consider the cycle $cyc=\left\lbrace 1,2,...,(m-1)\right\rbrace $ and construct a sub-matrix $\bm G_{sub}\in \mathbb{R}^{(m-1)\times (m-1)}$ from it. For $i-$th row of $\bm G_{sub}$,  we set it to be a sequence that starts with $i$ in  $cyc$ and ends until it reaches $(m-1)$ elements. For example, the first row will be $\left[ 1,2,...,(m-1)\right] $, the second row will be $\left[ 2,3,...,(m-1),1\right] $, and so on.
	Take the first $(m-1)$ rows and first $(m-1)$ columns of $\bm G$ to be $\bm G_{sub}$ and fill last row and last column of $\bm G$ with zeros.
	Then obviously $\bm G$ fulfills Conditions 2 and 3.

	To fulfill Condition 1, set  $\bm G_{i,m}$ and $ \bm G_{m,i}$ to be $ \bm G_{ii}$ and then set $\bm G_{ii} = 0$ for $i=1,...,(m-1)$. By this operation, it's easy to see that for first $(m-1)$ rows  and  first $(m-1)$ columns, they are still permutations of sequence $\left\lbrace 0,1,2,...,(m-1)\right\rbrace $ and symmetrization of $\bm G$ is not violated. It remains to prove that last row and last column are also a permutation of the sequence, which is equivalent to proving the diagonal part of $\bm G_{sub}$ is a permutation. In fact
	$\bm G_{sub(i,i)}$ is $(2i-1)$-th element of cycle $cyc$, $i=1,2,...,(m-1)$.  Since $m$ is even, diagonal parts of $\bm G_{sub}$ will cover the whole sequence  $\left\lbrace1,2,...,(m-1) \right\rbrace $.

	So for every pair $(j,j')$, $1\leq j <  j'\leq m$, we can assign it to Group $G_k$ where $k=\bm G_{j.j'}$. In this way, we decompose the collection $\left\lbrace (j,j'): 1\leq j <  j'\leq m\right\rbrace $ into $(m-1)$ groups where each group contains $m/2$ elements and within one group, there is no repeated individual.
\end{proof}

\begin{lemma}
	\label{regularization bound}
	Under the same assumptions as Theorem 2,  if $\lambda = c_{\lambda} \xi_n^2 $ with some constant $c_\lambda > 2\czhat $, then there exists a constant  $R_2$ depending on $I(\C_0)$, $\czhat$ and $c_\lambda$,
	such that with probability at least $1-\exp(\uni\frac{n}{\log n}\xi_{n,m}^2)$, we have
	$$ I(\hat \C) \leq R_2.$$
\end{lemma}

\begin{proof}
	From the basic inequality \eqref{basic_inequality_convex}, we have
	\begin{align}
	\|\Delta\|_{n,m}^2 + \lambda I (\hat \C) & \leq  2 \langle \error, \Delta \rangle +  \lambda I (\C_0),  \\
	\label{eqn:reg}
	\lambda I (\hat \C) & \leq  2 \langle \error, \Delta \rangle +  \lambda I (\C_0).
	\end{align}
	From Theorem 2, we know that
	\begin{align}
	\label{eqn:inner}
	\langle \error, \Delta\rangle\leq    \czhat \xi_{n,m}^2 \left\lbrace I(\hat \C )+ I( \C_0)\right\rbrace  + 2\czhat \xi_{n,m}  \|\Delta\|_{n,m},
	\end{align}
	and
	\begin{align}
	\label{eqn:final}
	\|\Delta\|_{n,m} \leq \left\lbrace 2c_\lambda I(\C_0)\right\rbrace ^{\frac{1}{2}}\xi_{n,m} +  4\czhat \xi_{n,m}.
	\end{align}

	Therefore, plug  \eqref{eqn:final}  into \eqref{eqn:final},
	\begin{align*}
	\langle \error, \Delta\rangle\leq \left[  \czhat\left\lbrace  I(\hat \C ) + I( \C_0)\right\rbrace  + 2\czhat \left\lbrace 2c_\lambda I(\C_0)\right\rbrace ^{\frac{1}{2}} + 8\czhat^2\right] \xi_{n,m}^2.
	\end{align*}
	By plugging in\eqref{eqn:reg}, we have
	\begin{align*}
	(c_\lambda-2\czhat) I(\hat \C) \leq 2\czhat I(\C_0) + 4 \czhat \left\lbrace 2c_\lambda I(\C_0)\right\rbrace ^{\frac{1}{2}} + 16 \czhat^2 + c_\lambda I(\C_0).
	\end{align*}
	Therefore,  there exists a constant $L_2$, such that
	$$ I(\hat \C) \leq \frac{2\czhat I(\C_0) + 4 \czhat \left\lbrace 2c_\lambda I(\C_0)\right\rbrace ^{\frac{1}{2}} + 16 \czhat^2 + c_\lambda I(\C_0)}{c_\lambda-2\czhat} \leq R_2.$$
\end{proof}

\begin{lemma}
	\label{tensor_sequence}
	Suppose $K_1(\cdot,\cdot)= K_2(\cdot,\cdot)= \ldots K_p (\cdot,\cdot)$, then $\Hscr_1=\Hscr_2=\ldots=\Hscr_p$. If eigenvalues of $K_k$ has decay $\mu_n^{(k)} \asymp (n^{-s})$ for some constant s. Then eigenvalues of the reproducing kernel for  tensor product $\bigotimes_{k=1}^p \Hscr_k \otimes \bigotimes_{k=1}^p \Hscr_k$ will have decay $\mu_n \asymp (n^{-s} (\log n)^{s(2p-1)})$
\end{lemma}

\begin{proof}
	A direct application of Theorem 1 \citep{krieg2018tensor} completes the proof. 
\end{proof}

\begin{lemma}
	\label{order}
	Take $t$ to be the solution of the equality $$\frac{1}{\sqrt{nm}} \left(  \sum_{h=1}^{\infty} \min\left\lbrace t^2,  \mu_h\right\rbrace \right)^{1/2}  = t^2,$$
	where $\mu_h\asymp  (h^{-2\alpha} (\log l)^{2\alpha(2p-1)})$. Then as $n\rightarrow \infty$ and $m\rightarrow\infty$, the solution $$t \asymp (nm)^{-\frac{\alpha }{1+2\alpha }}(\log nm)^{\frac{\alpha(2p-1)}{2\alpha +1 }}.$$
\end{lemma}

\begin{proof}
	Take $N = nm$, To find the order of $t$. We need to find $l'$ such that $t^2 \asymp l'^{-2\alpha} (\log l') ^{2\alpha (2p-1)}$. From some simple analysis, we could see that when $N\rightarrow \infty$, $t \rightarrow 0$ and $l'\rightarrow \infty$. Therefore, when $N\rightarrow \infty$ we have
	\begin{align*}
	t &\asymp l'^{-\alpha} (\log l') ^{\alpha (2p-1)},\\
	\frac{1}{t} &\asymp l'^{\alpha} (\log l') ^{-\alpha (2p-1)},\\
	log(1/t) &\asymp \alpha \log l' - \alpha(2p-1) \log(\log(l'))  \asymp \log l',\\
	l'&\asymp t^{-1/\alpha} (\log(1/t))^ {2p-1}.
	\end{align*}

	It's easy to see that $t^2 l' \asymp (t)^{2-1/\alpha}(\log (1/t))^{2p-1}$,
	$\sum_{l\ge l'} \mu_l \asymp \bigO (l'^{-2\alpha +1} (\log l')^{2\alpha(2p-1)}) \asymp \bigO (t^{2-1/\alpha}(\log(1/t))^{2p-1})$.

	So  $\sum_{l\ge l'} \mu_l  =\bigO (\xi_n^2 l' )$, therefore
	\begin{align*}
	\frac{\sqrt{t^2 l'}}{\sqrt{N}} &\asymp t ^2,\\
	N&\asymp (1/t)^{2+1/\alpha} (\log (1/ t)) ^ {2p-1}, \\
	\log N &\asymp (2+1/\alpha) log(1/ \xi_n) + (2p-1) \log\log(1/t) \asymp \log (1/t),\\
	1/t&\asymp  N^{\frac{\alpha }{1+2\alpha }} (\log N)^{-\frac{\alpha(2p-1)}{2\alpha +1 }},\\
	t &\asymp  N^{-\frac{\alpha }{1+2\alpha }} (\log N)^{\frac{\alpha(2p-1)}{2\alpha +1 }}.
	\end{align*}

\end{proof}

\section{Simulation}

\subsection{Eigenfunctions in different simulation settings}
\label{supp:eigen}
We present three simulation settings in a table form (Tables \ref{set1}, \ref{set2} and \ref{set3}). In each table, rows correspond to basis functions for dimension 1 and columns correspond to basis functions for dimension 2. Recall that for each dimension, we use $e_k(t) = \sqrt{2} \cos (k\pi t)$, $k = 1,2,\dots$ as basis.  Then, for the cell with position row $i$ and column $j$, it represents the two dimensional function $f_{ij}(s_1,s_2) = e_i(s_1)e_j(s_2)$.  We use a positive integer $k$ to indicate that this two dimensional function is the $k$-th eigenfunction.
The details of the three settings are given as follows.

\begin{table}[ht]
	\centering
	\begin{tabular}{| l |  l  l |}
		\hline
		& $e_1$ & $e_2$ \\
		\hline
		$e_1$ & 1 &  2      \\
		$e_2$ &  3  & 5               \\
		$e_3$ &   4  & 6             \\
		\hline
	\end{tabular}
	\caption{Eigenfunctions for Setting 1 }
	\label{set1}
\end{table}

\begin{table}[ht]
	\centering
	\begin{tabular}{| l |  l l l  l |}
		\hline
		& $e_1$ & $e_2$ & $e_3$ & $e_4$ \\
		\hline
		$e_1$ & 1 &  2  & \_ & \_    \\
		$e_2$ &  3  & 4 & \_ & \_               \\
		$e_3$ &   \_  & \_ & 5 & \_            \\
		$e_4$ &  \_ &\_ &\_ &\_ 6\\
		\hline
	\end{tabular}
	\caption{Eigenfunctions for Setting 2 }
	\label{set2}
\end{table}

\begin{table}[ht]
	\centering
	\begin{tabular}{| l |  l l l  l |}
		\hline
		& $e_1$ & $e_2$ & $e_3$ & $e_4$ \\
		\hline
		$e_1$ & \_ &  1  & \_ & \_    \\
		$e_2$ &  2  & \_ & \_ & \_               \\
		$e_3$ &   \_  & \_ & 3 & \_            \\
		$e_4$ &  \_ &\_ &\_ &\_ 4\\
		\hline
	\end{tabular}
	\caption{Eigenfunctions for Setting 3 }
	\label{set3}
\end{table}

Setting 1:  $R=6$, $r_1=3$, $r_2=2$. For dimension 1, we use  $e_1$, $e_2$ and $e_3$ as our basis functions.
For dimension 2, we use $e_1$ and $e_2$ as our basis functions.
Let 6 eigenfunctions $\psi_k$ be the tensor product of these one dimensional basis with eigenvalue decay $\lambda_k = 1/(k^2)$, $k=1,2,...,6$.
Eigenfunctions can be expressed as $\psi_k (t_1,t_2) =e_i(t_1)e_j(t_2)$, where $k = 2(i-1) + j$ for $k = 1,2,3,6$ and $\psi_4(s,t) = e_3(s)e_1(t)$, $\psi_5(t_1,t_2)= e_2(t_1)e_2(t_2)$.
In this setting, $R = r_1 * r_2$, one-way basis are mostly shared among different eigenfunctions.

Setting 2: $R = 6$, $r_1=r_2=4$. 	For both  dimension 1 and dimension 2, we use $e_i$, $i=1,..,4$  as our basis functions. Let 6 eigenfunctions $\psi_k$  with eigenvalue decay $\lambda_k = 1/(k^2)$, $k=1,2,...,6$.
$\psi_k (t_1,t_2) =e_i(t_1)e_j(t_2)$, where $k = 2(i-1) + j$ for $k = 1,2,3$.  $\psi_k(t_1,t_2) = e_{k-2}(t_1)e_{k-2}(t_2)$ for $k=4,5,6$.
In this setting,  one-way basis are partly shared by different eigenfunctions.

Setting 3: $R=r_1=r_2=4$.  	For both  dimension 1 and dimension 2, we use $e_i$, $i=1,..,4$  as our basis functions.
Let 4  eigenfunctions $\psi_k$ with  eigenvalue decay $\lambda_k = 1/(k^2)$, $k=1,...,4$.   $\psi_1(t_1,t_2)=e_1(t_1)e_2(t_2)$, $\psi_2(t_1,t_2)=e_2(t_1)e_1(t_2)$ and $\psi_k(t_1,t_2)=e_k(t_1)e_k(t_2)$ for $k=3,4$.
In this case, one-way basis are not shared among different eigenfunctions.

\subsection{Additional simulation results for sparse design}
\label{supp:ran_n100}
The simulation results for the sparse design with sample size $n=100$ are shown in Table \ref{ran_n100}.

\begin{table}[ht]
	\centering
	\resizebox{\textwidth}{!}{%
		{\small
			\begin{tabular}{crr | l llll}
				\hline
				Setting & $m$ & $\sigma$ &  & $\convex$ & $\Wong$ & \llsm  & \llsmp \\
				\hline
				1 & 10 & 0.1 & AISE & 0.101 (5.47e-03) & 0.122 (1.20e-02) & 4.36 (2.28e+00) & 1.702 (8.43e-01)\\
				&  &  & $\bar R$ & 7.66 & 2.45& - & 142.61 \\
				&  &  & $\bar r_1$, $\bar r_2$  & 5.07 , 5.04 & \_ & \_ &\_\\
				\hline
				&  & 0.4 & AISE & 0.104 (5.62e-03) & 0.12 (1.19e-02) & 3.89 (1.78e+00) & 0.989 (1.96e-01)\\
				&  &  & $\bar R$& 7.34 & 2.2& - & 146.33\\
				&  &  & $\bar r_1$, $\bar r_2$  & 4.84 , 4.82  & \_ & \_ &\_\\
				\hline
				& 20 & 0.1 & AISE & 0.0661 (2.99e-03) & 0.075 (3.49e-03) & 3.93 (3.17e+00) & 1.40 (9.80e-01)\\
				&  &  & $\bar R$ & 7.84 & 3.02  & - & 249.95\\
				&  &  &$\bar r_1$, $\bar r_2$  & 5.48 , 5.48 & \_ & \_ &\_\\
				\hline
				&  & 0.4 & AISE & 0.0679 (3.06e-03) & 0.0761 (3.34e-03) & 0.468 (6.90e-02) & 0.310 (2.32e-02)\\
				&  &  & $\bar R$ & 7.51  & 2.83 & - &205.675\\
				&  &  & $\bar r_1$, $\bar r_2$  & 5.38 , 5.38 & \_ & \_ &\_\\
				\hline
				2 & 10 & 0.1 & AISE & 0.1 (5.38e-03) & 0.113 (6.12e-03) & 2.12 (6.23e-01) & 0.826 (1.76e-01)\\
				&  &  & $\bar R$ & 7.68 & 2.38 & - &144.645\\
				&  &  & $\bar r_1$, $\bar r_2$& 5.53, 5.56 & \_ & \_ &\_\\
				\hline
				&  & 0.4 & AISE & 0.102 (5.44e-03) & 0.112 (5.64e-03) & 4.18 (2.21e+00) & 0.931 (1.76e-01)\\
				&  &  & $\bar R$& 7.34 & 2.22 & - & 146.855 \\
				&  &  &$\bar r_1$, $\bar r_2$& 5.49, 5.49 & \_ & \_ &\_\\
				\hline
				& 20 & 0.1 & AISE & 0.0637 (2.95e-03) & 0.0706 (3.21e-03) & 0.472 (8.01e-02) & 0.304 (2.80e-02)\\
				&  &  &$\bar R$ & 8.37 & 2.76 & - & 200.69\\
				&  &  &$\bar r_1$, $\bar r_2$& 5.81, 5.8 & \_ & \_ &\_\\
				\hline
				&  & 0.4 & AISE & 0.0649 (3.06e-03) & 0.0733 (3.30e-03) & 0.484 (7.27e-02) & 0.317 (2.53e-02)\\
				&  &  & $\bar R$ & 8.24 & 2.78 & - & 206.16\\
				&  &  & $\bar r_1$, $\bar r_2$& 5.78, 5.78 & \_ & \_ &\_\\
				\hline
				3 & 10 & 0.1 & AISE & 0.105 (4.75e-03) & 0.115 (7.58e-03) & 24.1 (2.28e+01) & 1.87 (1.19)\\
				&  &  & $\bar R$ & 8.75 & 2.82 & - & 150.8\\
				&  &  & $\bar r_1$, $\bar r_2$ & 5.26, 5.32 & \_ & \_ & \_\\
				\hline
				&  & 0.4 & AISE & 0.11 (4.96e-03) & 0.115 (8.33e-03) & 26.2 (2.40e+01) & 2.05 \\
				&  &  & $\bar R$ & 9.44 & 2.74 & -&152.575\\
				&  &  &  $\bar r_1$, $\bar r_2$ & 5.37, 5.4 & \_ & \_& \_  \\
				\hline
				& 20& 0.1& AISE & 0.0698 (2.74e-03) & 0.0813 (4.63e-03) & 0.614 (2.28e-01) & 0.350 (8.35e-02)\\
				&  &  & $\bar R$ & 6.63 & 3.24 & - & 210.515\\
				&  &  & $\bar r_1$, $\bar r_2$ & 5.08, 5.14 & \_ & \_ & \_ \\
				\hline
				&  & 0.4 & AISE & 0.0721 (2.89e-03) & 0.0859 (5.03e-03) & 0.573 (1.74e-01) & 0.344 (6.37e-02)\\
				&  &  & $\bar R$ & 6.74 & 3.38 & - & 214.455 \\
				&  &  &  $\bar r_1$, $\bar r_2$& 5.11, 5.21 & \_ & \_ & \_ \\
				\hline
	\end{tabular}}}
	\caption{Simulation results for three Settings with the sparse design when sample size is 100 ($n=100$): see description in Table 1.}
	\label{ran_n100}
\end{table}

\subsection{Simulation results for regular design}
\label{supp:dense}
\renewcommand{\arraystretch}{1}
For regular design, we selected $10$ equally spaced points for each dimension and constructed a regular $10 \times 10$ grid ($m=100$). We set sample size  to be $50$ ($n=50$). Two different noise levels are considered, since regular design has dense observations, we pick $\sigma = 0.4$ to represent the low noise level and $\sigma=0.8$ to represent the high noise level. Beside methods we mentioned in sparse design, we also include an additional estimator from \cite{wang2017regularized} (\spatpca), which allows to perform multi-dimensional covariance function estimation with location-fixed observations into our comparisons. Results are showed  in Table \ref{grid_set1}, Table \ref{grid_set2} and Table \ref{grid_set3}.

\begin{table}[ht]
	\centering
	\resizebox{\textwidth}{!}{
		\begin{tabular}{rl | lllll}
			\hline
			$\sigma$ &  & $\convex$ & $\Wong$ & \llsm & \spatpca & \llsmp \\
			\hline
			0.40 & AISE & 0.0611 (4.37e-03) & 0.0626 (4.10e-03) & 0.0571 (3.71e-03) & 0.0625 (4.37e-03) &0.057 (3.71e-03)\\
			& $\hat R$ & 8.27 & 7.25 & - & 5.95 & 15.125 (0.10)\\
			& $\hat r_1, \hat r_2$ & 6, 6 & \_ & \_ & \_ & \_ \\
			\hline
			0.80 & AISE & 0.0629 (4.45e-03) & 0.0676 (4.49e-03) & 0.0643 (3.79e-03) & 0.0738 (4.52e-03) & 0.0639 (3.79e-03) \\
			& $\hat R$ & 10.9 & 3.98 & - & 5.84 &26.065 (0.087)\\
			& $r_1, r_2$ & 6, 6 & \_ & \_ & \_ & \_ \\
			\hline
	\end{tabular}}
	\caption{Results for Setting 1 on regular design: see description in Table 1.}
	\label{grid_set1}
\end{table}

\begin{table}[ht]
	\centering
	\resizebox{\textwidth}{!}{%
		\begin{tabular}{rl | lllll}
			\hline
			$\sigma$ &  & $\convex$ & $\Wong$ & \llsm & \spatpca& \llsmp \\
			\hline
			0.40 & AISE & 0.0602 (4.38e-03) & 0.0641 (4.69e-03) & 0.056 (3.71e-03) & 0.0624 (4.37e-03)  & 0.0559 (3.71e-03)\\
			& $\hat R$ & 8.09 & 7.22 & - & 4.21 &14.135 (0.095) \\
			& $\hat r_1, \hat r_2$ & 6, 6 & \_ & \_ & \_ & \_\\
			\hline
			0.80 & AISE & 0.062 (4.48e-03) & 0.0659 (4.54e-03) & 0.0631 (3.79e-03) & 0.0724 (4.47e-03)  & 0.0627 (3.79e-03)\\
			& $\hat R$ & 10.7 & 4.04 & - & 4.28  &25.84 (0.0898)\\
			&$\hat r_1, \hat r_2$ & 6, 6 & \_ & \_ & \_ & \_\\
			\hline
	\end{tabular}}
	\caption{Results for Setting 2 on regular design: see description in Table 1.}
	\label{grid_set2}
\end{table}

\begin{table}[ht]
	\centering
	\resizebox{\textwidth}{!}{%
		\begin{tabular}{rl | lllll}
			\hline
			$\sigma$ &  & $\convex$ & $\Wong$ & \llsm & \spatpca  & \llsmp \\
			\hline
			0.40 & AISE & 0.0628 (4.22e-03) & 0.0589 (4.34e-03) & 0.0677 (3.92e-03) & 0.0598 (4.17e-03)  &  0.0675 (3.92e-03)\\
			&$\hat R$ & 5.66 & 14 & - & 3.52 & 18.74 (0.104)\\
			& $\hat r_1, \hat r_2$ & 6, 6 & \_ & \_ & \_ & \_ \\
			\hline
			0.80 & AISE & 0.0645 (4.05e-03) & 0.0677 (4.48e-03) & 0.0745 (3.94e-03) & 0.0715 (4.20e-03)  & 0.07389 (3.94e-03)\\
			& $\hat R$  & 7.7 & 13.1 & - & 2.93 &29.485 (0.143)\\
			& $\hat r_1, \hat r_2$ & 6, 6 & \_ & \_ & \_ & \_ \\
			\hline
	\end{tabular}}
	\caption{Results for Setting 3 on regular design: see description in Table 1.}
	\label{grid_set3}
\end{table}

\bibliographystyle{chicago}
\bibliography{supplementary-plain.bbl}

\begin{table}[ht]
	\centering
	\resizebox{\textwidth}{!}{%
		{\small
			\begin{tabular}{rrr | l llll}
				\hline
				$n$ & $m$ & $\sigma$ &  & $\convex$ & $\Wong$ & \llsm  & \llsmp \\
				\hline
				100 & 10 & 0.1 & AISE & 0.101 (5.47e-03) & 0.122 (1.20e-02) & 4.36 (2.28e+00) & 1.702 (8.43e-01)\\
				&  &  & $\bar R$ & 7.66 & 2.45& - & 142.61 \\
				&  &  & $\bar r_1$, $\bar r_2$  & 5.07 , 5.04 & \_ & \_ &\_\\
				\hline
				&  & 0.4 & AISE & 0.104 (5.62e-03) & 0.12 (1.19e-02) & 3.89 (1.78e+00) & 0.989 (1.96e-01)\\
				&  &  & $\bar R$& 7.34 & 2.2& - & 146.33\\
				&  &  & $\bar r_1$, $\bar r_2$  & 4.84 , 4.82  & \_ & \_ &\_\\
				\hline
				& 20 & 0.1 & AISE & 0.0661 (2.99e-03) & 0.075 (3.49e-03) & 3.93 (3.17e+00) & 1.40 (9.80e-01)\\
				&  &  & $\bar R$ & 7.84 & 3.02  & - & 249.95\\
				&  &  &$\bar r_1$, $\bar r_2$  & 5.48 , 5.48 & \_ & \_ &\_\\
				\hline
				&  & 0.4 & AISE & 0.0679 (3.06e-03) & 0.0761 (3.34e-03) & 0.468 (6.90e-02) & 0.310 (2.32e-02)\\
				&  &  & $\bar R$ & 7.51  & 2.83 & - &205.675\\
				&  &  & $\bar r_1$, $\bar r_2$  & 5.38 , 5.38 & \_ & \_ &\_\\
				\hline
				200 & 10 & 0.1 & AISE & 0.053 (1.97e-03) & 0.0632 (3.22e-03) & 0.652 (1.92e-01) & 0.337 (5.35e-02)\\
				&  &  & $\bar R$& 8.38 & 2.94 & - &172.70 \\
				&  &  &$\bar r_1$, $\bar r_2$  & 5.4, 5.4& \_ & \_&\_ \\
				\hline
				&  & 0.4 & AISE & 0.0547 (2.01e-03) & 0.0656 (2.72e-03) & 0.714 (2.11e-01) &0.366 (5.96e-02)\\
				&  &  & $\bar R$& 9.16 & 2.84  & - &177.3 \\
				&  &  & $\bar r_1$, $\bar r_2$  & 5.34, 5.32 & \_ & \_ &\_\\
				\hline
				& 20 & 0.1 & AISE & 0.0343 (1.46e-03) & 0.0421 (1.97e-03) & 0.297 (1.39e-02) & 0.206 (4.62e-03)\\
				&  &  & $\bar R$ & 8.38 & 3.78 & - &317.44\\
				&  &  & $\bar r_1$, $\bar r_2$  & 5.84, 5.82 & \_ & \_ &\_\\
				\hline
				&  & 0.4 & AISE & 0.0354 (1.52e-03) & 0.044 (2.21e-03) & 0.325 (1.58e-02) & 0.223 (4.94e-03)\\
				&  &  & $\bar R$ & 8.86 & 3.76 & - & 326.31\\
				&  &  & $\bar r_1$, $\bar r_2$  & 5.83, 5.84& \_ & \_ &\_\\
				\hline
	\end{tabular}}}
	\caption{Simulation results for Setting 1 ($R=6$, $r_1=3$, and $r_2=2$) with the sparse design. The AISE values with standard errors (SE) in parentheses are provided for the four covariance estimators in comparison, together with average two-way ranks ($\bar R$) for those estimators which can lead to rank reduction (i.e., \convex, \Wong, and \llsmp) and average one-way ranks ($r_1$, $r_2$) for $\convex$.
		\label{ran_set1}
	}
\end{table}

\begin{table}[ht]
	\centering
	\resizebox{\textwidth}{!}{
		{\small
			\begin{tabular}{rrr | l llll}
				\hline
				$n$ & $m$ & $\sigma$ &  & $\convex$ & $\Wong$ & \llsm & \llsmp \\
				\hline
				100 & 10 & 0.1 & AISE & 0.1 (5.38e-03) & 0.113 (6.12e-03) & 2.12 (6.23e-01) & 0.826 (1.76e-01)\\
				&  &  & $\bar R$ & 7.68 & 2.38 & - &144.645\\
				&  &  & $\bar r_1$, $\bar r_2$& 5.53, 5.56 & \_ & \_ &\_\\
				\hline
				&  & 0.4 & AISE & 0.102 (5.44e-03) & 0.112 (5.64e-03) & 4.18 (2.21e+00) & 0.931 (1.76e-01)\\
				&  &  & $\bar R$& 7.34 & 2.22 & - & 146.855 \\
				&  &  &$\bar r_1$, $\bar r_2$& 5.49, 5.49 & \_ & \_ &\_\\
				\hline
				& 20 & 0.1 & AISE & 0.0637 (2.95e-03) & 0.0706 (3.21e-03) & 0.472 (8.01e-02) & 0.304 (2.80e-02)\\
				&  &  &$\bar R$ & 8.37 & 2.76 & - & 200.69\\
				&  &  &$\bar r_1$, $\bar r_2$& 5.81, 5.8 & \_ & \_ &\_\\
				\hline
				&  & 0.4 & AISE & 0.0649 (3.06e-03) & 0.0733 (3.30e-03) & 0.484 (7.27e-02) & 0.317 (2.53e-02)\\
				&  &  & $\bar R$ & 8.24 & 2.78 & - & 206.16\\
				&  &  & $\bar r_1$, $\bar r_2$& 5.78, 5.78 & \_ & \_ &\_\\
				\hline
				200 & 10 & 0.1 & AISE & 0.0532 (1.98e-03) & 0.0636 (3.12e-03) & 2.33 (1.13e+00) &0.795 (2.98e-01)\\
				&  &  & $\bar R$ & 8.48 & 3.02 & - &191.175 \\
				&  &  & $\bar r_1$, $\bar r_2$& 5.82, 5.82 & \_ & \_ &\_\\
				\hline
				&  & 0.4 & AISE & 0.0548 (2.05e-03) & 0.0686 (3.53e-03) & 2.44 (1.17e+00) & 0.828 (3.04e-01)\\
				&  &  & $\bar R$ & 9.04 & 3.04 & - & 196.34\\
				&  &  & $\bar r_1$, $\bar r_2$ & 5.71, 5.74 & \_ & \_ &\_\\
				\hline
				& 20 & 0.1 & AISE & 0.0341 (1.43e-03) & 0.0419 (2.02e-03) & 0.301 (1.58e-02) &0.208 (4.50e-03)\\
				&  &  & $\bar R$ & 8.99 & 3.74 & - & 318.645\\
				&  &  & $\bar r_1$, $\bar r_2$ & 5.93, 5.92 & \_ & \_ &\_\\
				\hline
				&  & 0.4 & AISE & 0.0348 (1.43e-03) & 0.043 (2.22e-03) & 0.328 (1.78e-02) & 0.225 (4.74e-03)\\
				&  &  &$\bar R$& 8.01 & 3.6 & - & 327.395\\
				&  &  &$\bar r_1$, $\bar r_2$& 5.94, 5.93 & \_ & \_ &\_ \\
				\hline
	\end{tabular}}}
	\caption{Simulation results for Setting 2 ($R=6$ and $r_1=r_2=4$) with the sparse design: see description in Table \ref{ran_set1}.}
	\label{ran_set2}
\end{table}

\begin{table}[ht]
	\centering
	\caption{Simulation results for Setting 3 ($R=r_1=r_2=4$) with the sparse design: see description in Table \ref{ran_set1}.}
	\resizebox{\textwidth}{!}{
		{\small
			\begin{tabular}{rrr | lllll}
				\hline
				$n$ & $m$ & $\sigma$ &  & $\convex$ & $\Wong$ & \llsm & \llsmp \\
				\hline
				100 & 10 & 0.1 & AISE & 0.105 (4.75e-03) & 0.115 (7.58e-03) & 24.1 (2.28e+01) & 1.87 (1.19)\\
				&  &  & $\bar R$ & 8.75 & 2.82 & - & 150.8\\
				&  &  & $\bar r_1$, $\bar r_2$ & 5.26, 5.32 & \_ & \_ & \_\\
				\hline
				&  & 0.4 & AISE & 0.11 (4.96e-03) & 0.115 (8.33e-03) & 26.2 (2.40e+01) & 2.05 \\
				&  &  & $\bar R$ & 9.44 & 2.74 & -&152.575\\
				&  &  &  $\bar r_1$, $\bar r_2$ & 5.37, 5.4 & \_ & \_& \_  \\
				\hline
				& 20& 0.1& AISE & 0.0698 (2.74e-03) & 0.0813 (4.63e-03) & 0.614 (2.28e-01) & 0.350 (8.35e-02)\\
				&  &  & $\bar R$ & 6.63 & 3.24 & - & 210.515\\
				&  &  & $\bar r_1$, $\bar r_2$ & 5.08, 5.14 & \_ & \_ & \_ \\
				\hline
				&  & 0.4 & AISE & 0.0721 (2.89e-03) & 0.0859 (5.03e-03) & 0.573 (1.74e-01) & 0.344 (6.37e-02)\\
				&  &  & $\bar R$ & 6.74 & 3.38 & - & 214.455 \\
				&  &  &  $\bar r_1$, $\bar r_2$& 5.11, 5.21 & \_ & \_ & \_ \\
				\hline
				200 & 10& 0.1 & AISE & 0.058 (2.62e-03) & 0.0692 (5.33e-03) & 0.454 (7.28e-02) & 0.286 (2.89e-02) \\
				&  &  & $\bar R$ & 6.26 & 3.12 & -& 182.74 \\
				&  &  &  $\bar r_1$, $\bar r_2$ & 5, 5.06 & \_ & \_ & \_ \\
				\hline
				&  & 0.4 & AISE & 0.0598 (2.68e-03) & 0.0733 (6.14e-03) & 0.531 (1.07e-01) &0.323 (4.23e-02)\\
				&  &  & $\bar R$ & 6.48 & 3.2 & - & 185.82 \\
				&  &  &  $\bar r_1$, $\bar r_2$ & 4.99, 5.07 & \_ & \_ & \_ \\
				\hline
				& 20& 0.1 & AISE & 0.0422 (1.37e-03) & 0.0535 (2.64e-03) & 0.267 (5.04e-03) & 0.196 (3.59e-03)\\
				&  &  & $\bar R$& 6.29 & 4.49 & - & 332.09 \\
				&  &  &  $\bar r_1$, $\bar r_2$ & 5.62, 5.69 & \_ & \_ & \_ \\
				\hline
				&  & 0.4 & AISE & 0.0424 (1.30e-03) & 0.0494 (2.42e-03) & 0.292 (5.30e-03) &0.212 (3.72e-03)\\
				&  &  & $\bar R$ & 5.68 & 3.36 & - &338.725 \\
				&  &  &  $\bar r_1$, $\bar r_2$ & 5.59, 5.66 & \_ & \_ & \_ \\
				\hline
	\end{tabular}}}
	\label{ran_set3}
\end{table}

\clearpage